\documentclass[12pt]{article}

\catcode`\@=11
\@addtoreset{equation}{section}

\global\arraycolsep=1pt

\usepackage{amsbsy,amssymb,latexsym,amsfonts, amsmath}
\usepackage{mathrsfs}
\usepackage{graphicx}

\RequirePackage[dvips,usenames]{color}
\definecolor{fireblick}{rgb}{0.698039,0.133333,0.133333}

\newcommand{\beq}{\begin{equation}}
\newcommand{\eeq}{\end{equation}}
\newcommand{\bea}{\begin{eqnarray}}
\newcommand{\eea}{\end{eqnarray}}

\newcommand{\CN}{{\mathcal N}}

\def\Tr{\mathop{\rm Tr}}

\newcommand{\im}{\mathrm{i}}
\newcommand{\ex}{\mathrm{e}}
\newcommand{\de}{\mathrm{d}}

\renewcommand{\thefootnote}{\fnsymbol{footnote}}

\begin{document}
%
%
\begin{titlepage}
\begin{flushright}
\normalsize
~~~~
OCU-PHYS 324\\
YITP-09-75\\
November, 2009 \\
revised April, 2010 \\
\end{flushright}

\vspace{15pt}

\begin{center}
{\LARGE
The Quiver Matrix Model} \\
\vspace{6pt}
{\LARGE and} \\
\vspace{6pt}
{\LARGE $2d-4d$ Conformal Connection}\\
\end{center}

\vspace{23pt}

\begin{center}
{ H. Itoyama$^{a, b}$\footnote{e-mail: itoyama@sci.osaka-cu.ac.jp},
K. Maruyoshi$^c$\footnote{e-mail: maruyosh@yukawa.kyoto-u.ac.jp},
and
T. Oota$^b$\footnote{e-mail: toota@sci.osaka-cu.ac.jp}
}\\
%
\vspace{18pt}
%

$^a$ \it Department of Mathematics and Physics, Graduate School of Science\\
Osaka City University\\
\vspace{5pt}

$^b$ \it Osaka City University Advanced Mathematical Institute (OCAMI)

\vspace{5pt}

3-3-138, Sugimoto, Sumiyoshi-ku, Osaka, 558-8585, Japan \\

\vspace{10pt}

$^c$ \it Yukawa Institute for Theoretical Physics, Kyoto University, Kyoto 606-8502, Japan\\

\end{center}
%
\vspace{20pt}
\begin{center}
Abstract\\
\end{center}
We review  the quiver matrix model (the ITEP model)
in the light of the recent progress on 2d-4d
connection of conformal field theories, in particular,
on the relation between Toda field theories and
a class of quiver superconformal gauge theories.
On the basis of the CFT representation of the $\beta$ deformation of the model,
a quantum spectral curve is introduced as
$ \langle \! \langle \det ( x- \im g_s \partial \phi(z))
\rangle\! \rangle=0$ at finite $N$ and
for $\beta \neq 1$.
The planar loop equation in the large $N$ limit follows with the aid of
$W_n$ constraints.
Residue analysis is provided both for the curve
of the matrix model with the ``multi-log''
potential and for the Seiberg-Witten curve in the case of $SU(n)$ with $2n$ flavors,
leading to the matching of the mass parameters. The isomorphism of the two curves is made manifest.


\vfill

\setcounter{footnote}{0}
\renewcommand{\thefootnote}{\arabic{footnote}}

\end{titlepage}

\renewcommand{\thefootnote}{\arabic{footnote}}
\setcounter{footnote}{0}

\section{Introduction}
\label{sec:intro}

Matrix models have had successes in several stages
of the developments in string theory, gauge theory
and related studies of integrable systems.  
A list of those in the last twenty years  include
2d gravity, exact evaluation of gluino condensate prepotential,  topological strings, etc.
 
  Recent progress has been triggered
by the construction of a large class of
$\mathcal{N}=2$ superconformal $SU(n)$ ``generalized quiver''
gauge theories in four dimensions by Gaiotto \cite{Gaiotto}.
(See also \cite{GM,Tachikawa:2009rb,Benini:2009gi,
Nanopoulos:2009xe,Drukker:2009tz}).
Subsequently an interesting conjecture has been made
by Alday, Gaiotto and Tachikawa (AGT) \cite{AGT}
(and its $SU(n)$ generalization by \cite{Wyllard})
on the equivalence of the Nekrasov partition function \cite{Nekrasov,NO,NY} and
the $2d$ conformal block of the Toda field theory.
These are followed by a number of extensive checks and
pieces of supporting evidence
\cite{Maruyoshi:2009uk,MMM0907,Gaiotto:2009hg,gai0908,
MMMM,MM0908a,MM0908b,NX0908,AGGTV,
DGOT,Benini:2009mz,MMM0909a,DV,MMM0909b,
pog,MM0909,BT,AY,ABT,
Papadodimas:2009eu,MM0910,AM0910,Gaiotto:2009fs,Nanopoulos:2009uw,
MM0911}.
Very recently, Dijkgraaf and Vafa \cite{DV} have suggested
an explanation of this phenomenon
by  the so-called quiver matrix model \cite{MMM,KMMMP,kos1,kos2,DV2002b}.
(In this paper, it is occasionally referred to as the ITEP model \cite{MMM,KMMMP}).
Their reasoning is based on
the matrix model realization of type IIB topological
strings on
a local Calabi-Yau with a local $A_{n-1}$ singularity 
which geometrically engineers the gauge theory.
By choosing ``multi-Penner'' potentials \cite{pen} for the $A_{n-1}$
quiver matrix model, they argued that
the spectral curve of the matrix model
at large $N$ (the size of the matrix)
can be understood as the Seiberg-Witten curve
of the attendant $SU(n)$ generalized quiver superconformal gauge theory (SCFT).

    The AGT conjecture is  regarded as a more concrete realization of
the folklore connection between $4d$  gauge theories
and the attendant $2d$ sigma models
(see \cite{pol} for instance),
that led  to the study of two dimensional
quantum integrable field theories in late seventies
(\cite{tha} for instance).
 In the light of  potential importance of this subject,
we find it useful to devote  the substantial part of
the present paper in  reviewing  and reformulating the basic structure
of the quiver matrix model \cite{IMNO}.
 
 The quiver matrix model
associated with Lie algebra $\mathfrak{g}$
of ADE type with rank $r$ is obtained
as a solution to extended Virasoro constraints, i.e., $W(\mathfrak{g})$
constraints \cite{FZ,FL,BS}
at finite $N$ \cite{MMM,KMMMP}\footnote{
This kind of reasoning is seen
in the construction of other eigenvalue models \cite{AlvarezGaume:1991jd}.}.
This fact that the model automatically implements the  $W(\mathfrak{g})$
constraints at finite $N$  is an advantage over more traditional
two- and multi-matrix models
where  finite $N$ Schwinger-Dyson equations are 
typically $w_{1+\infty}$ type and are more involved
\cite{Itoyama:1991hz,deBoer:1991qw,Ahn:1992ay,Bonora:1993ie}.
The model is defined by using $r$ independent
free massless chiral bosons in two dimensional CFT
with the central charge $c=r$ and the final form of the partition
function is formulated as the
integrations over eigenvalues of the matrices.
 A key ingredient of its construction is a set of
 screening charges of the $c=r$ CFT.
By construction,
the partition function respects the $W(\mathfrak{g})$,
the extended Casimir algebra
generated by higher spin currents which commute with the screening
charges.
The partition function can be reformulated as
integrations over the matrices in the adjoint and bi-fundamental
representations.

In the context of the AGT conjecture for more general Toda CFTs,
it is natural to consider ``$\beta$-ensembles'' of ADE quiver matrix models.
In this paper, using the $\beta$-ensemble of
the $A_{n-1}$ quiver matrix model at finite $N$,
we define a non-commutative Calabi-Yau threefold
with the quantum deformed $A_{n-1}$ singularity as
\beq
u v + \bigl\langle \!\! \bigl\langle
\det( x - \im g_s \partial \phi(z) )
\bigr\rangle \!\!  \bigr\rangle = 0, \qquad
[ x, z] = -\im Q g_s.
\eeq
For $n=3$, it takes the following form
\beq
\begin{split}
u v &+ \bigl( x - t_3(z) \bigr) \bigl( x - t_2(z) \bigr)
\bigl( x - t_1(z) \bigr) \cr
& - \bigl( x - t_3(z) \bigr) f_1(z)
- f_2(z) \bigl( x - t_1(z) \bigr) - g_1(z) + g_2(z) = 0,
\end{split}
\eeq
with the non-commutativity $[x,z] = -\im Q g_s = \epsilon_1 + \epsilon_2$.

 In order to bring these analyses in the more recent context, 
we take a simple example, namely,
$\mathcal{N}=2$ $SU(3)$ SQCD with $6$ flavors to
provide a residue analysis of the quiver matrix model curve 
and that of the Seiberg-Witten curve
from type IIA and M-theory consideration.
This leads to the matching of the mass parameters.

A main result of this paper is to establish the isomorphism of the
spectral curve of the quiver matrix model ($A_n$ type and beta deformed in general) 
in the planar limit and the
 corresponding Seiberg-Witten curve in the Witten-Gaiotto form. This has become possible 
as we have reformulated the matrix model curve, starting from the $W_n$ constraints at finite $N$, 
proposing the curve as a form of the characteristic equation, and finally using the singlet 
factorization in the planar limit to write it in a form where the isomorphism is rather manifest. 
To the best of our knowledge, such a systematic investigation has not been attempted 
 and only clumsy expressions in the planar limit have been available. (See references in subsequent sections.)

This paper is organized as follows.
In section $2$, after giving several punchlines,
we review  known facts on
the one-matrix model, using the CFT notation and the Virasoro constraints at finite $N$ 
\cite{VirN}.
 In section $3$, ordinary ADE quiver matrix models are
reviewed in the same spirit and their ``$\beta$-ensemble'' is introduced.
For the case of $A_{n-1}$ quiver matrix model,
we show that the quantum spectral curve
\beq
\bigl\langle \!\! \bigl\langle
\det( x - \im g_s \partial \phi(z) )
\bigr\rangle \!\!  \bigr\rangle = 0, \qquad
[ x, z] = -\im Q g_s,
\eeq
is well-defined within the matrix model integral.
The partition function obeys the $W_n$ constraints at finite $N$
which are the properties of the original matrix integrals. 
The planar loop equation
follows from these structures together with the large $N$ factorization. 
In section $4$, we adopt Dijkgraaf-Vafa's recipe to
treat $\mathcal{N}=2$ superconformal gauge theory by using the quiver matrix model
with Penner like action.
We concentrate on the $A_2$ quiver matrix model and consider the spectral curve.
Section $5$ treats the Seiberg-Witten curve for
$SU(n)$ gauge theory with $N_f=2n$ flavors.
We see that this curve enjoys the same properties as those of the spectral curve.
In particular, we give a matching of the mass parameters  first for the case of $n=3$.
In order to render the isomorphism of the two curves clearer,  a subsection is devoted to
establishing this point. A matching of the mass parameters for general $n$ is readily given.

 The major reason that we emphasized starting with the finite $N$ spectral curve is already stated above: 
through this procedure and the large $N$ factorization of the singlet operators, the isomorphism
has been established in this paper. The continuous flow, beginning with the construction at finite $N$, 
and ending with the singlet factorization in the planar limit has been indispensable in order for this
paper to be legible and self-contained.

\section{One-matrix model and the $\beta$-ensemble}

In the case of $A_1$, the quiver matrix model corresponds to the Hermitian
one-matrix model \cite{BIPZ} (see also \cite{HMM,meh}).
The associated CFT is a single free boson with $c=1$.
On the other hand, the Liouville CFT which appears in AGT conjecture
for $SU(2)$ has the central charge $c= 1 + 6 Q^2$ with $Q=b+(1/b)$.
It is known that there is a one-matrix model which has a connection
with the CFT with $c=1+6Q^2$. It is the $\beta$-ensemble of one-matrix
model with $\beta= - b^2$.
It is easy to deal with the Liouville CFT by introducing the
Feign-Fuchs background charge in the CFT notation. The CFT notation works well
for the $\beta$-ensemble of the one-matrix model
and the appearance of the CFT with $c=1+6Q^2$
in the matrix model is a built-in result.
The case of $\beta=1$
corresponds to the ordinary Hermitian one-matrix model.
The AGT conjecture implies that these deformation parameters
are related to  Nekrasov's deformation parameters $\epsilon_1$
and $\epsilon_2$ by \cite{DV} \footnote{In AGT \cite{AGT},
a different parametrization with $\epsilon_1 = b$ and $\epsilon_2=1/b$
is used.}
\beq
\label{NekrasovCFT}
\epsilon_1 = - \im b g_s, \qquad \epsilon_2 = - \frac{\im g_s}{b}.
\eeq
Note that $\epsilon_1 \epsilon_2 = - g_s^2$ and 
$\epsilon_1/\epsilon_2 = b^2$.

It is known that the $\beta$-ensemble of one-matrix model
at finite $N$
is related to a quantum (non-commutative) spectral curve of the form
\beq
\left( x + \frac{1}{2} W'(z) \right)
\left( x - \frac{1}{2} W'(z) \right) - f(z) = 0,
\eeq
whose non-commutativity is given by
\beq
[ x, z] = - \im Q g_s  = \epsilon_1 + \epsilon_2.
\eeq
Using the $\beta$-ensemble matrix model, it is possible to define
a non-commutative local Calabi-Yau threefold
with a quantum deformed $A_1$ singularity
\beq
u v + \left( x + \frac{1}{2} W'(z) \right)
\left( x - \frac{1}{2} W'(z) \right) - f(z) = 0.
\eeq
It can be written as the form
\beq
uv + x^2 - g^2 \langle \! \langle T(z) \rangle \! \rangle = 0,
\eeq
where $\langle\! \langle \ \ \rangle\! \rangle$
is one-matrix model average over the $\beta$-ensemble
and $T(z)$ is the energy momentum
tensor of $c=1+6Q^2$ CFT expressed as the collective field
of the matrix eigenvalues. Later, we will give exact definitions
of these quantities.

\subsection{Hermitian one-matrix model: undeformed case}

In this case the relevant matrix model 
is the Hermitian one-matrix model.
The partition function takes the form
\beq
\label{HMMZ}
Z = \int [ \de M] \exp \left( \frac{1}{g_s} \mathrm{Tr}\, 
W(M) \right).
\eeq
Here $M$ is an $N \times N$ Hermitian matrix.
Correlation function of this matrix model is defined by
\beq
\label{CFMM}
\langle \! \langle \mathcal{O} \rangle \! \rangle
:= \frac{1}{Z} \int [ \de M ] \, \mathcal{O} \exp
\left( \frac{1}{g_s} \mathrm{Tr} \, W(M)  \right).
\eeq
Here $\mathcal{O} = \mathcal{O}(M)$ is a function of the Hermitian matrix.

The partition function \eqref{HMMZ} can be written in terms
of the eigenvalues $\lambda_I$ of the matrix $M$:
\beq
Z = \int \de^N \lambda \, \Delta(\lambda)^2 
\exp\left( \sum_{I=1}^N \frac{1}{g_s} W(\lambda_I) \right),
\eeq
where $\Delta(\lambda)$ is the Vandermonde determinant
\beq
\Delta(\lambda) = \prod_{1 \leq I <  J \leq N}
( \lambda_I - \lambda_J ).
\eeq

It is well-known that the Hermitian matrix model has a close connection
with the $c=1$ free chiral boson. The partition function
can be rewritten in terms of CFT operators.
The mode expansion of the chiral boson $\phi(z)$ is chosen as follows:
\beq
\phi(z) = \phi_0 - \im a_0 \log z + \im \sum_{ n \neq 0}
\frac{a_n}{n} z^{-n},
\eeq
and the non-trivial commutation relations are given by 
\beq
[ a_n, a_m ] = n \delta_{n+m,0}, \qquad
[ \phi_0, a_0 ] = \im.
\eeq
Hence, our normalization of the correlator is given by
\beq
\langle \phi(z) \phi(w) \rangle = - \log(z-w).
\eeq
The energy momentum tensor with the central charge $c=1$ is given by
\beq
\label{EMT}
T(z) = - \frac{1}{2} : \bigl( \partial \phi(z) \bigr)^2 :
= \sum_{ n \in \mathbb{Z}} L_n z^{-n-2}.
\eeq
The screening charges which commute with the Virasoro generators $L_n$
are given by
\beq
Q_{\pm} = \int \de z : \ex^{\pm \im \sqrt{2} \phi(z)} :,
\eeq
with a certain integration contour.

The Fock vacuum is given by
\beq
a_n | 0 \rangle = 0, \qquad
\langle 0 | a_{-n} = 0, \qquad
n \geq 0.
\eeq
Let
\beq
\langle N |:= \langle 0 | \ex^{- \im \sqrt{2} N \phi_0}.
\eeq
Then, the partition function \eqref{HMMZ} of the Hermitian matrix
model can be rewritten in terms of the free chiral boson as follows
\beq
\label{HMMZ2}
Z = \langle N | \exp\left( \frac{1}{2 \sqrt{2} \pi g_s}
\oint \de z\,  W(z) \partial \phi(z) \right)
(Q_+)^N | 0 \rangle.
\eeq
Associated with this expression, for an operator $\mathcal{O}$
constructed from the boson oscillators, we use the following notation
\beq
\langle \mathcal{O} \rangle_{\mathrm{CFT}}:=
\frac{1}{Z} \, \langle N | 
\exp\left( \frac{1}{2 \sqrt{2} \pi g_s}
\oint \de z\,  W(z) \partial \phi(z) \right)
\, \mathcal{O}\, (Q_+)^N | 0 \rangle.
\eeq
Within the normal ordering, correlators consisting of the
chiral boson $\phi(z)$ in CFT
have their counterparts in the matrix model correlators \eqref{CFMM}:
\beq
\bigl\langle : \mathcal{O}(\phi) \dotsm :  \bigr\rangle_{CFT}
= \bigl\langle \! \bigl\langle \, 
\mathcal{O}(\phi) \dotsm \bigr\rangle \! \bigr\rangle.
\eeq
In the matrix model correlator, the chiral boson is
realized as a collective field of the eigenvalues:
\beq
\im \phi(z) = \frac{1}{\sqrt{2} g_s } W(z) + \sqrt{2}\,  
\mathrm{Tr} \, \mathrm{Log}(z - M).
\eeq

It is known that the partition function
of the Hermitian matrix model at finite $N$ 
obeys the Virasoro constraints \cite{VirN}.
In the CFT language, it follows from the commutativity of the
Virasoro generators $L_n$ with the screening charge $Q_+$
and
\beq
L_n | 0 \rangle = 0, \qquad n \geq -1.
\eeq
In the ITEP construction, the Virasoro constraints manifestly hold at finite $N$ as  
\beq
\bigl \langle L_n \bigr \rangle_{\mathrm{CFT}} = 0, \qquad
n \geq -1.
\eeq
They are equivalent to the regularity of the correlator of the energy momentum tensor
\beq
\bigl\langle  T(z) \bigr\rangle_{CFT} = 
\bigl \langle \! \bigl \langle
T(z) \bigr \rangle \! \bigr \rangle 
\eeq
at $z=0$.

Now, using the help of the Hermitian matrix model correlator,
a local Calabi-Yau threefold with $A_1$
singularity over a Riemann surface $\Sigma$
can be defined by
\beq
uv + x^2 - g_s^2 \, 
\bigl \langle \! \bigl \langle
T(z) \bigr \rangle \! \bigr \rangle = 0.
\eeq

Using the collective field expression
\beq
T(z) = - \frac{1}{2} \bigl(\partial \phi(z))^2, \qquad
\im \partial \phi(z) = \frac{1}{\sqrt{2} g_s} W'(z)
+ \sqrt{2} \, \mathrm{Tr}\, \frac{1}{z - M},
\eeq
we have (for the derivation, see the next subsection, below \eqref{EM1})   
\beq
g_s^2 \bigl \langle \! \bigl \langle
T(z) \bigr \rangle \! \bigr \rangle 
= \frac{1}{4} W'(z)^2 + f(z),
\eeq
where
\beq
f(z) = \left\langle \!\!\! \left\langle
g_s \sum_{I=1}^N \frac{W'(z) - W'(\lambda_I)}{z - \lambda_I}
\right\rangle \!\!\! \right\rangle.
\eeq

Hence, the local Calabi-Yau threefold is a surface
in $(u,v,x,z) \in \mathbb{C}^4$ defined by
\beq
u v + x^2 - \frac{1}{4} W'(z)^2 - f(z) = 0.
\eeq
At $uv=0$, it describes some algebraic curve $\Sigma$
in $(x,z) \in \mathbb{C}^2$:
\beq
x^2 - \frac{1}{4} W'(z)^2 - f(z) = 0.
\eeq
Note that this algebraic curve is well-defined for finite $N$
due to the Virasoro constraints of the matrix model.

\subsection{$\beta$-ensemble}

Nekrasov's deformation corresponds to the modification 
of the energy-momentum tensor \eqref{EMT} by the introduction of 
the background charge \`{a} la Feign-Fuchs:
\beq
\label{EMT2}
T(z) = - \frac{1}{2} : \partial \phi(z)^2 : + \frac{Q}{\sqrt{2}}
\partial^2 \phi(z), \qquad
Q = b + \frac{1}{b}.
\eeq
This energy momentum tensor has the central charge $c=1+6 Q^2$.
Undeformed case is recovered  
at $b=\im$, $Q=0$ and $\epsilon_1 = - \epsilon_2 = g_s$.

Screening charges for this energy momentum tensor are given by
\beq
Q_+(b) = \int \de z\, : \ex^{\sqrt{2} b \phi(z) }:, \qquad
Q_-(b) = \int \de z\, : \ex^{\sqrt{2} b^{-1} \phi(z) } :. 
\eeq
Let
\beq
\langle N; b|:= \langle
0 | \ex^{ - \sqrt{2} b N \phi_0}.
\eeq
It is natural to consider the following deformation of
the partition function \eqref{HMMZ2}:
\beq
\label{bMMZ}
\begin{split}
Z&:= \langle N; b |
\exp\left( \frac{1}{2 \sqrt{2} \pi g_s}
\oint \de z W(z) \partial \phi(z) \right)
\Bigl( Q_+(b) \Bigr)^N | 0 \rangle \cr
&= \int \de^N \lambda \,
\bigl(\Delta(\lambda)\bigr)^{-2b^2}
\exp\left( - \frac{\im b}{g_s} \sum_{I=1}^N W(\lambda_I)
\right).
\end{split}
\eeq
This matrix model is known as the $\beta$-ensemble
\cite{meh,har,EM,bet} with $\sqrt{\beta} = -\im b$.
For $\beta=1/2, 1, 2$, it corresponds to the integrations
over an orthogonal, hermitian and symplectic matrix respectively.

Instead of the screening charge $Q_+(b)$, we can use $Q_-(b)$
to express the partition of the $\beta$-ensemble model. 
The corresponding expressions are obtained by replacing $b$ with
$b^{-1}$.

It is known that this partition function is related to
a \textit{non-commutative} 
(or \textit{quantum}) \textit{spectral curve} \cite{EM}.
In this case, the non-commutativity is given by
\beq
[ x, z] = - \im Q g_s = - \im \left( b + \frac{1}{b} \right) g_s 
= \epsilon_1 + \epsilon_2.
\eeq
For $Q \neq 0$, $x$ can be realized as a differential operator
$x = - \im Q g_s \partial/\partial z$.

Note that the energy-momentum tensor \eqref{EMT2}
can be defined by
the Miura transformation:
\beq
: 
\left( x + \frac{\im g_s}{\sqrt{2}} \partial \phi(z)\right)
\left( x - \frac{\im g_s}{\sqrt{2}} \partial \phi(z) \right):
= x^2 - g_s^2 T(z).
\eeq

The collective field expression of the chiral boson $\phi(z)$
now becomes
\beq
\im \phi(z) = \frac{1}{\sqrt{2} g_s} W(z) - \im b \sqrt{2}\,
\mathrm{Tr} \, \mathrm{Log}( z - M ).
\eeq
Note that
\beq
J(z) = \im \partial \phi(z)
= \frac{1}{\sqrt{2} g_s} W'(z) - \im b  \sqrt{2}\, 
\mathrm{Tr} \, \frac{1}{z-M}.
\eeq
Using the collective field expression, we see that
\beq
\label{EM1}
\begin{split}
g_s^2 T(z) &= \frac{1}{4} (W'(z))^2 - \frac{\im}{2} Q \, g_s W''(z)
- \im b g_s \sum_{I=1}^N
\frac{W'(z) - W'(\lambda_I)}{z - \lambda_I} \cr
& - \im b g_s \sum_{I=1}^N \frac{W'(\lambda_I)}{z- \lambda_I}
+ \sum_{I=1}^N \frac{g_s^2}{(z-\lambda_I)^2}
- 2 b^2 g_s^2 \sum_{I=1}^N \frac{1}{z-\lambda_I}
\sum_{J \neq I} \frac{1}{\lambda_I - \lambda_J}.
\end{split}
\eeq
The terms in the second line of \eqref{EM1} correspond 
to the ``singular'' part\footnote{Later, we consider the case in which
$W(z)$ has logarithmic singularities.}
of the energy momentum tensor and we can see that 
\beq
T(z) \Bigr|_{\mathrm{singular \ part}} \times  \ex^{-S_{\mathrm{eff}}}
= \sum_{I=1}^N
\frac{\partial}{\partial \lambda_I}
\left( \frac{1}{z - \lambda_I}
\ex^{-S_{\mathrm{eff}}} \right),
\eeq
where
\beq
\ex^{-S_{\mathrm{eff}}}=
\bigl(\Delta(\lambda)\bigr)^{-2b^2} 
\exp\left( - \frac{\im b}{g_s} \sum_{I=1}^N W(\lambda_s)
\right). 
\eeq

Therefore the Virasoro constraints for the deformed one-matrix model 
imply that
\beq
\langle \! \langle g_s^2 T(z) \rangle\! \rangle
= \frac{1}{4} W'(z)^2 - \frac{\im}{2} Q \, g_s W''(z)
+ f(z),
\eeq
where
\beq
f(z):= \left\langle \!\! \!\left\langle
 - \im b g_s \sum_{I=1}^N \frac{W'(z) - W'(\lambda_I)}{z- \lambda_I} 
\right\rangle \!\! \!\right \rangle.
\eeq
Here the matrix model average is defined as in the undeformed case. Explicitly
for some function $\mathcal{O}(\lambda)$ of the eigenvalues, we have
\beq
\langle\! \langle \mathcal{O}(\lambda) \rangle \! \rangle
= \frac{1}{Z}
\int \de^N \lambda \bigl( \Delta(\lambda) \bigr)^{-2b^2}
\mathcal{O}(\lambda) 
\exp\left( - \frac{\im b}{g_s} \sum_{I=1}^N W(\lambda_I) \right).
\eeq

Hence the quantum spectral curve related to the $\beta$-ensemble
is defined by
\beq
\label{QSC}
\left\langle \!\! \left\langle
\left( x + \frac{\im g_s}{\sqrt{2}} \partial \phi(z) \right)
\left( x - \frac{\im g_s}{\sqrt{2}} \partial \phi(z) \right)
\right\rangle \! \! \right \rangle =
x^2 - g_s^2 \langle \! \langle T(z) \rangle \! \rangle =  0.
\eeq
Explicitly, it is given by
\beq
x^2 - \frac{1}{4} W'(z)^2 + \frac{\im}{2} Q \, g_s W''(z)
- f(z) = 0.
\eeq
Note that this quantum spectral curve can be rewritten as
\beq
\left( x + \frac{1}{2} W'(z) \right)
\left( x - \frac{1}{2} W'(z) \right) - f(z) = 0.
\eeq

Therefore, the associated local Calabi-Yau threefold also
becomes a non-commutative surface:
\beq
u v + \left( x + \frac{1}{2} W'(z) \right)
\left( x - \frac{1}{2} W'(z) \right) - f(z) = 0, \qquad
[ x, z] = - \im Q g_s.
\eeq
Strictly speaking, for $Q \neq 0$, $x$ is a differential operator
and the quantum spectral curve is a differential equation for some
``wave function.'' This equation has a close connection with
$W$-gravity and Hitchin systems. For recent discussion on this point
in the light of the AGT conjecture, 
see \cite{Gaiotto:2009hg,BT,Nanopoulos:2009uw}
and references therein. From the point of view of 
the $\mathcal{D}$-module, see \cite{DHS}.
In string theory, the non-commutativity corresponds to
turn on a constant NS two-form.

Our main concern is the large $N$ 
string/gauge duality.
In the planar limit (a large $N$ limit with the
't Hooft coupling $S = g_s N$ kept finite),
$g_s \rightarrow 0$ and thus
the non-commutativity vanishes\footnote{The Nekrasov function in 
a different limit such that $\epsilon_2\rightarrow 0$ 
while keeping $\epsilon_1$ finite is investigated in
\cite{NS,MM0910,MM0911}. In this limit, the non-commutativity 
remains finite.}
: $[x,z] \rightarrow 0$.

\subsection{Large $N$ limit}

The partition function \eqref{bMMZ}
has a topological expansion
\beq
\label{TE}
Z = \exp\left( \sum_{k=0}^{\infty} g_s^{k-2} \mathcal{F}_{k/2} \right).
\eeq

In the large $N$ limit with the 't Hooft coupling $S = g_s N$
kept finite, leading contribution comes from the planar part 
$\mathcal{F}_0$ in \eqref{TE} and can be evaluated by the saddle
point method.
For simplicity, we assume that the parameter $b$
is pure imaginary and $\beta = - b^2 > 0$. In this case
Nekrasov's deformation parameters $\epsilon_1$
and $\epsilon_2$ are real.

The stationary conditions $\partial S_{\mathrm{eff}}/\partial \lambda_I=0$
yield
\beq
\label{stationary}
W'(\lambda_I) - 2 \im b g_s 
\sum_{\stackrel{\scriptstyle J=1}{(J \neq I)}}^N
 \frac{1}{\lambda_I- \lambda_J} = 0, \qquad
I=1,2,\dotsc, N.
\eeq
Since we assume that $b$ is pure imaginary,
these stationary equations have real solutions $\lambda_I$. 
We evaluate the partition function around 
a classical solution with certain filling fractions $\nu_i=N_i/N$
around the local extrema $W'(a_i)=0$.

In the planar limit,
the large $N$ factorization yields
\beq
\lim_{g_s \rightarrow 0} 
\left\langle \!\! \left\langle  
\left( x + \frac{\im g_s}{\sqrt{2}} \partial \phi(z) \right)
\left( x - \frac{\im g_s}{\sqrt{2}} \partial \phi(z) \right)
\right\rangle \!\! \right\rangle
= (x + y(z)) (x - y(z))= x^2 - y(z)^2, 
\eeq
where
\beq
\label{yz}
y(z) := \lim_{g_s \rightarrow 0} \frac{ g_s}{\sqrt{2}}  
\langle \! \langle \im \partial \phi(z) \rangle \! \rangle
=\frac{1}{2} W'(z) - \im b S \int 
\frac{\rho(\lambda)}{z - \lambda} \de \lambda.
\eeq
Here $\rho(\lambda)$ is the density function of the solution
to the stationary conditions \eqref{stationary}.
\beq
\rho(\lambda) = \lim_{N \rightarrow \infty}
\frac{1}{N} \sum_{I=1}^N \delta( \lambda - \lambda_I).
\eeq
The stationary conditions in the large $N$ limit go to
\beq
W'(\lambda) - 2 \im b S \, 
\mathrm{P}\! \int \frac{\rho(\lambda')}{\lambda- \lambda'} \de \lambda'=0.
\eeq
Here $\mathrm{P}$ denotes the principal value. Note that in the stationary
conditions and in the definition of $y(z)$,
the parameter $b$ always appears in the combination 
$\tilde{S} = - \im b S$.
Therefore, if we replace the 't Hooft coupling $S = g_s N$
by $\tilde{S} = -\im b S = -\im b g_s N$, we can use the large $N$ formulas
of the undeformed Hermitian one-matrix model. We will call $\tilde{S}$
\textit{deformed 't Hooft coupling}.

Hence, in the large $N$ limit, the local Calabi-Yau is deformation
of $A_1 \times \mathbb{C}$
\beq
uv + x^2 - y^2(z) = 0,
\eeq
and the algebraic curve $\Sigma$ becomes
\beq
x^2 - y^2(z) = 0,
\eeq
and the points $(x,z)$ on $\Sigma \subset \mathbb{C}^2$
can be covered by two sheets $(x,z)
= ( \pm y(z), z )$.

In the large $N$ limit, the Virasoro constraints become an algebraic
equation
\beq
y^2(z) = \frac{1}{4} W'(z)^2 + f(z).
\eeq

\section{ADE quiver matrix models and their ``$\beta$-ensemble''}
\label{sec:matrix}

In this section, we first briefly review the ADE quiver matrix models 
\cite{MMM,KMMMP,kos1,kos2,DV2002b}.
An excellent review for the undeformed case can be found in \cite{CKR}.
Then, we introduce the ``$\beta$-ensemble'' or deformed ADE quiver matrix models.
For $A_{n-1}$ cases, they can be found in \cite{AMOS1}.

Using the $\beta$-deformed $A_{n-1}$ quiver matrix model,
we introduce a non-commutative local Calabi-Yau threefold related to
deformations of $A_{n-1}$ singularities.

\subsection{ADE quiver matrix models and CFT with $c=r$}

Let $\mathfrak{g}$ be a finite dimensional 
Lie algebra of ADE type with rank $r$,
$\mathfrak{h}$ the Cartan subalgebra of $\mathfrak{g}$,
and $\mathfrak{h}^*$ its dual. 
We sometimes denote the natural pairings
between $\mathfrak{h}$ and $\mathfrak{h}^*$ by $\langle \cdot, \cdot \rangle$:
\beq
\alpha(h) = \langle \alpha, h \rangle, \qquad
\alpha \in \mathfrak{h}^*, \ \ h \in \mathfrak{h}.
\eeq
Let $\alpha_a \in \mathfrak{h}^*$ 
$(a=1,2,\dotsc, r)$ be simple roots of $\mathfrak{g}$
and $( \cdot, \cdot)$ is the inner product on $\mathfrak{h}^*$.
Our normalization is chosen as $(\alpha_a, \alpha_a)=2$.
The fundamental weights are denoted by
$\Lambda^a$ $(a=1,2,\dotsc, r)$
\beq
( \Lambda^a, \alpha_b^{\vee} ) = \delta^a_b, \qquad
\alpha_a^{\vee} = \frac{2 \alpha_a}{(\alpha_a, \alpha_a)}.
\eeq

In the Dynkin diagram of $\mathfrak{g}$
we associate $N_a \times N_a$ Hermitian matrices $M_a$
with vertices $a$ for simple roots $\alpha_a$,
and complex $N_a \times N_b$ matrices $Q_{ab}$ and their
Hermitian conjugate $Q_{ba} = Q_{ab}^{\dag}$ with links
connecting vertices $a$ and $b$. We label links of the Dynkin
diagram by pairs of node label $(a,b)$ with an ordering $a<b$.
Let $\mathcal{E}$ and $\mathcal{A}$ be the set of ``edges'' $(a,b)$
(with $a<b$) and the set of ``arrows'' $(a,b)$ respectively:
\beq
\mathcal{E}:= \{ (a,b) \, | \, 1 \leq a<b \leq r, \ 
(\alpha_a, \alpha_b) = -1 \},
\eeq
\beq
\mathcal{A}:= \{ (a, b) \, | \, 1 \leq a,b \leq r, \ (\alpha_a, \alpha_b)
= - 1 \}.
\eeq
  
The partition function of the 
quiver matrix model 
\cite{MMM,KMMMP,kos1,kos2,DV2002b} associated with $\mathfrak{g}$
is given by
\beq
Z = \int \prod_{a=1}^r [ \de M_a ] \prod_{(a,b) \in
\mathcal{A} } [ \de Q_{ab} ]
\exp\left( \frac{1}{g_s} W(M, Q) \right),
\eeq
where
\beq
W(M, Q) = \im \sum_{(a,b) \in \mathcal{A}} 
s_{ab} \mathrm{Tr} \, Q_{ba} M_a Q_{ab}
+ \sum_{a=1}^r \mathrm{Tr}\, W_a( M_a),
\eeq
with real constants $s_{ab}$ obeying the conditions $s_{ab} = - s_{ba}$. 
Note that
\beq
\prod_{(a,b) \in \mathcal{A}}[ \de Q_{ab} ]
= \prod_{(a,b) \in \mathcal{E}} [ \de Q_{ba} \de Q_{ab} ],
\eeq
\beq
\sum_{(a,b) \in \mathcal{A}}
s_{ab} \mathrm{Tr} \, Q_{ba} M_a Q_{ab}
= \sum_{(a,b) \in \mathcal{E}}
s_{ab} \bigl( \mathrm{Tr} \, Q_{ba} M_a Q_{ab} - \mathrm{Tr}\, 
Q_{ab} M_b Q_{ba} \bigr).
\eeq
The integration measures $[\de M_a ]$ and $[ \de Q_{ba} \de Q_{ab} ]$
are defined by using the metrics 
$\mathrm{Tr} ( \de M_a)^2$ and $\mathrm{Tr}( \de Q_{ba} \de Q_{ab})$
respectively.

Integrations over $Q_{ab}$ are Gauss-Fresnel type and
are easily performed:
\beq
\label{GF}
\int [ \de Q_{ba} \de Q_{ab} ]
\exp\left( \frac{\im s_{ab}}{g_s}
\bigl( \mathrm{Tr} \, Q_{ba} M_a Q_{ab} - \mathrm{Tr}\, 
Q_{ab} M_b Q_{ba} \bigr) \right)
= \det\bigl( M_a \otimes 1_{N_b} - 1_{N_a} \otimes M_b^T 
\bigr)^{-1},
\eeq
where $1_n$ is the $n \times n$ identity matrix and $T$
denotes transposition. For simplicity 
we have chosen the normalization
of the measure $[\de Q_{ba} \de Q_{ab}]$ to set the proportional
constant in the right-handed side of \eqref{GF} to be unity.  

Now the integrand depends only on the eigenvalues of $r$ Hermitian
matrices $M_a$.
Let us denote them by $\lambda^{(a)}_I$ ($a=1,2,\dotsc, r$
and $I=1,2,\dotsc, N_a$).
The partition function of the quiver matrix model (ITEP model)
now reduces to the form of integrations
over the eigenvalues of $M_a$ \cite{MMM,KMMMP}:
\beq 
\label{part}
Z = \int \prod_{a=1}^r \left\{ \prod_{I=1}^{N_a} \de \lambda^{(a)}_I \right\}\, 
\Delta_{\mathfrak{g}}(\lambda)
\exp\left( \sum_{a=1}^r \sum_{I=1}^{N_a} \frac{1}{g_s} W_a(\lambda^{(a)}_I)
\right),
\eeq
where $W_a$ is a potential and
\beq
\Delta_{\mathfrak{g}}(\lambda) = 
\prod_{a=1}^r \prod_{1 \leq I < J \leq N_a}
( \lambda_I^{(a)} - \lambda_J^{(a)})^2
\prod_{1 \leq a< b \leq r}
\prod_{I=1}^{N_a} \prod_{J=1}^{N_b}
( \lambda^{(a)}_I - \lambda^{(b)}_J )^{(\alpha_a, \alpha_b)}.
\eeq

The partition function \eqref{part}
can be rewritten in terms of CFT operators.
Let $\phi(z)$ be $\mathfrak{h}$-valued massless chiral field
and $\phi_a(z):= \langle \alpha_a, \phi(z) \rangle$. Their correlators 
are given by
\beq
\langle \phi_a(z) \phi_b(w) \rangle = - ( \alpha_a, \alpha_b ) \log(z-w),
\qquad
a,b=1,2,\dotsc, r.
\eeq
The modes
\beq
\phi(z) = \phi_0 - \im a_0 \log z + \im \sum_{n \neq 0} 
\frac{a_n}{n} z^{-n} \in \mathfrak{h}
\eeq
obey the commutation relations
\beq
[ \langle \alpha, a_n \rangle, \langle \beta, a_m \rangle ]
= n \delta_{n+m,0} ( \alpha, \beta), \qquad
[ \langle \alpha, \phi_0 \rangle, \langle \beta, a_0 \rangle]
= \im ( \alpha, \beta), \qquad \alpha, \beta \in \mathfrak{h}^*.
\eeq
The Fock vacuum is given by
\beq
\alpha(a_n)|0 \rangle = 0, \qquad
\langle 0 | \alpha(a_{-n}) = 0, \qquad
n \geq 0, \qquad
\alpha \in \mathfrak{h}^*.
\eeq
Let
\beq
\langle \{ N_a \} |:= \langle 0|
\exp\left( - \im \sum_{a=1}^r N_a \alpha_a( \phi_0) \right).
\eeq
It is convenient to introduce the $\mathfrak{h}^*$-valued potential $W(z)$
by
\beq
W(z):= \sum_{a=1}^r W_a(z) \Lambda^a \in \mathfrak{h}^*.
\eeq
Note that $W_a(z) = ( \alpha^{\vee}_a, W(z) )$.

The energy-momentum tensor is given by
\beq
T(z) = - \frac{1}{2} : \mathcal{K}\bigl( \partial \phi(z), \partial \phi(z) \bigr):,
\eeq
where $\mathcal{K}$ is the Killing form. Let $H^i$ ($i=1,2,\dotsc, r$)
be an orthonormal basis of the Cartan subalgebra $\mathfrak{h}$
with respect to the Killing form:
$\mathcal{K}(H^i, H^j) = \delta^{ij}$. In this basis, 
the components of the $\mathfrak{h}$-valued chiral boson are
just $r$ independent free chiral bosons: 
\beq
\phi(z) = \sum_{i=1}^r H^i \phi_i(z), \qquad
\langle \phi_i(z) \phi_j(w) \rangle = - \delta_{ij} \log(z-w),
\eeq
and the energy-momentum tensor in this basis is given by
\beq
T(z) = - \frac{1}{2} \sum_{i=1}^r : \bigl( \partial \phi_i(z) \bigr)^2:.
\eeq
The central charge is $c=r$.

Note that for a root $\alpha$, $[H^i, E_{\alpha}] = \alpha^i E_{\alpha}$
with $\alpha^i = \alpha(H^i) = \langle \alpha, H^i \rangle$.
Then, the bosons $\phi_a(z)$ associated with the simple roots $\alpha_a$
are expressed in this basis as follows:
\beq
\phi_a(z) = \langle \alpha_a, \phi(z) \rangle
= \sum_{i=1}^r \alpha_a^i \phi_i(z) \equiv \alpha_a \cdot \phi(z),
\qquad a=1,2,\dotsc, r.
\eeq
For roots $\alpha$ and $\beta$, 
the inner product on the root space is expressed in their components as
$( \alpha, \beta) = \sum_{i=1}^r \alpha^i \beta^i$. Here
$\alpha^i = \alpha(H^i)$ and
$\beta^i = \beta(H^i)$.

The screening charges associated with the simple roots are defined by
\beq
Q_a:= \int \de z \, : \ex^{\im \phi_a(z)}:, \qquad a=1,2,\dotsc, r,
\eeq
with an appropriate contour integration.

Using these definitions, the partition function \eqref{part}
can be written as follows
\beq
Z = \langle \{ N_a \} | \, 
\exp\left( \frac{1}{2\pi g_s} \oint_{\infty}
\de z \, \langle W(z), \partial \phi(z) \rangle \right) 
( Q_1)^{N_1} \dotsm (Q_r)^{N_r} \, | 0 \rangle.
\eeq

The chiral scalar field appears in the matrix model
as the collective field of the eigenvalues
\beq
\label{coll}
\im \langle \alpha, \phi(z) \rangle
= ( \alpha, \frac{1}{g_s} W(z) ) 
+ \sum_{a=1}^r ( \alpha, \alpha_a) \log \det( z - M_a ),
\qquad \alpha \in \mathfrak{h}^*.
\eeq
In particular, for $\alpha=\alpha_a$,
\beq
\im \phi_a(z) = \frac{1}{g_s} W_a(z) + \sum_{b=1}^r
( \alpha_a, \alpha_b) \log \det( z - M_b ).
\eeq

\subsection{$\beta$-ensemble of ADE quiver matrix model}

Inspired by the recent AGT conjecture, we are interested in 
the conformal Toda field theory based on a finite-dimensional
Lie algebra $\mathfrak{g}$ of ADE type. 
For the conformal Toda field theories,
the energy momentum tensor is given by
\beq
T(z) = - \frac{1}{2} : \mathcal{K}
\bigl(\partial \phi(z), \partial \phi(z) \bigr):
+ Q \langle \rho, \partial^2 \phi(z) \rangle,
\eeq
where $Q = b + (1/b)$ and $\rho$ is the Weyl vector of $\mathfrak{g}$,
half the sum of the positive roots.
In the orthonormal basis, it takes the form
\beq
T(z) = - \frac{1}{2} \sum_{i=1}^r : 
\bigl( \partial \phi_i(z) \bigr)^2:
+ Q \sum_{i=1}^r \rho^i \partial^2 \phi_i(z)
= - \frac{1}{2} : \partial \phi(z) \cdot \partial \phi(z):
+ Q \rho \cdot \partial^2 \phi(z).
\eeq
The central charge is given by
\beq
\label{ccADE}
c = r + 12 Q^2 ( \rho, \rho) = r \Bigl\{ 1 + h ( h+1) Q^2 \Bigr\}.
\eeq
Here $h$ is the Coxeter number of the simply-laced Lie algebra
$\mathfrak{g}$ whose rank is $r$. 
Explicitly,
$h_{A_{n-1}} = n$ (with $r=n-1$), $h_{D_r} =2r-2$, $h_{E_6}=12$,
$h_{E_7}=18$ and $h_{E_8}= 30$.

The partition function of the
corresponding $\beta$-ensemble quiver matrix model
(with $\beta = -b^2$)
is the following deformation 
of the quiver matrix model \eqref{part}
\beq
\label{partb}
Z:=\int \prod_{a=1}^r 
\left\{ \prod_{I=1}^{N_a} \de \lambda^{(a)}_I \right\}
\Bigl(\Delta_{\mathfrak{g}}(\lambda) \Bigr)^{-b^2}
\exp\left( - \frac{\im b}{g_s} \sum_{a=1}^r \sum_{I=1}^{N_a}
W_a(\lambda^{(a)}_I) \right).
\eeq
At $b = \im$, it reduces to the original quiver matrix model \eqref{part}.

The corresponding collective field realization of 
chiral scalars is given by
\beq
\label{betaCF}
\im \langle \alpha, \phi(z) \rangle
= ( \alpha, \frac{1}{g_s} W(z) )
- \im b \sum_{a=1}^r ( \alpha, \alpha_a) \log \det( z - M_a),
\qquad \alpha \in \mathfrak{h}^*.
\eeq

Now we require the non-commutativity 
\beq
[ x, z ] = - \im Q \, g_s, \qquad
Q = b + \frac{1}{b}.
\eeq
If $Q \neq 0$, $x$ can be realized as $x = - \im Q \, g_s \partial$.
Here $\partial = \partial/\partial z$.

Note that in the simple root basis $\phi_a(z) = \alpha_a \cdot \phi(z)$,
the collective fields are given by
\beq
\label{srb}
\im \phi_a(z) = \frac{1}{g_s} W_a(z)
- \im b \sum_{a'=1}^r C_{aa'} \log ( z - M_a), \qquad
C_{aa'}:= ( \alpha_a, \alpha_{a'} ).
\eeq
The energy momentum tensor in this basis has the form
\beq
\label{EMTg}
T(z) = \sum_{a,a'=1}^r (C^{-1})^{aa'} \left(
- \frac{1}{2} : \partial \phi_a(z) \partial \phi_{a'}(z):
+ Q \, \partial^2 \phi_{a'}(z) \right).
\eeq
Here we have used $\rho = \sum_{a=1}^r \Lambda^a$.
Using \eqref{srb} and \eqref{EMTg}, we 
can check that the partition function \eqref{partb}
obey the Virasoro constraints.

\subsection{$A_{n-1}$ quiver matrix model and non-commutative spectral curve}

In this subsection, we consider 
the case of $\mathfrak{g} = A_{n-1} = \mathfrak{su}(n)$
with its rank $r=n-1$.
The generators of $\mathfrak{su}(n)$ algebra in the defining
representation are $n \times n$ traceless Hermitian matrices.
The generators of the Cartan subalgebra $\mathfrak{h}$ can be chosen
as diagonal ones. Using the defining representation, the Killing form
can be chosen as the trace of $n \times n$ matrices: $\mathcal{K}(X,Y)
= \mathrm{Tr}(XY)$ for $X, Y \in \mathfrak{g}$.
 
In the case of $A_{n-1}$ quiver matrix model,
it is convenient to denote the chiral boson as follows:
\beq
\label{Ancb}
\phi(z) = \mathrm{diag}(\varphi_1(z), \varphi_2(z), \dotsc, \varphi_n(z))
\in \mathfrak{h}, \qquad
\mathrm{Tr} \, \phi(z) = 0.
\eeq
Let $\varepsilon_i$ be a linear map from $n \times n$ diagonal
matrices to $\mathbb{C}$ such that 
\beq
\varepsilon_i\bigl( \mathrm{diag}(x_1, x_2, \dotsc, x_n) \bigr) = x_i,
\qquad i=1,2,\dotsc, n.
\eeq
The simple roots $\alpha_a$ and the fundamental weights $\Lambda^a$
of $A_{n-1}$ algebra are given by
\beq
\alpha_a = \varepsilon_a - \varepsilon_{a+1}, \qquad
\Lambda^a = \sum_{i=1}^a \varepsilon_i - \frac{a}{n} \sum_{i=1}^n 
\varepsilon_i, \qquad
a=1,2,\dotsc, n-1.
\eeq
With $(\varepsilon_i, \varepsilon_j) = \delta_{ij}$, we have
the (symmetrized) Cartan matrix of $A_{n-1}$:
\beq
C_{aa'} 
= (\alpha_a, \alpha_{a'} ) 
= 2 \delta_{aa'} - \delta_{a,a'+1} - \delta_{a+1,a'}.
\eeq
It follows that
$(\Lambda^a, \Lambda^{a'}) = (C^{-1})^{aa'} 
= (1/n)\, \mathrm{min}(a,a')
\{ n - \mathrm{max}(a,a') \}$.
The explicit form of the Weyl vector is given by 
\beq
\label{WVAn}
\rho = \frac{1}{2} \sum_{\alpha > 0} \alpha
= \frac{1}{2} \sum_{i=1}^n ( n - 2 i + 1) \varepsilon_i.
\eeq
Let $v_i \in \mathfrak{h}^*$ 
($i=1, 2,\dotsc, n$) be the $n$ weights in the
defining representation with the highest
weight $\Lambda^1$:
\beq
\label{nwt}
v_i = \Lambda^1 - \sum_{a=1}^{i-1} \alpha_a=
\varepsilon_i - \frac{1}{n} \sum_{j=1}^n \varepsilon_j,
\qquad i=1,2,\dotsc, n.
\eeq
Note that $(v_i, v_j) = \delta_{ij} - (1/n)$.
Using these weights, the components of the chiral boson \eqref{Ancb} 
can be obtained by $\varphi_i(z) = \langle v_i, \phi(z) \rangle$.
The relation between the $n-1$ bosons $\phi_a(z)$ associated with the simple roots
$\alpha_a$ and these $n$ bosons $\varphi_i(z)$ are given by
$\phi_a(z) = \langle \alpha_a, \phi(z) \rangle
= \varphi_a(z) - \varphi_{a+1}(z)$.
Their correlators are given by
\beq
\langle \phi_a(z) \phi_{a'}(w) \rangle = -( \alpha_a, \alpha_{a'})
\log(z-w), \qquad
a,a'=1,2,\dotsc, n-1,
\eeq
\beq
\langle \varphi_i(z) \varphi_j(w) \rangle
= - ( v_i, v_j) \log(z-w) 
= - \left( \delta_{ij} - \frac{1}{n} \right) \log(z-w), \qquad
i,j=1,2,\dotsc, n.
\eeq

Let us introduce $n$ spin-$1$ currents with one constraint as follows
\beq
\label{Ji}
J_i(z):= \im \partial \varphi_i(z), \qquad
i=1,2,\dotsc, n,
\qquad
\sum_{i=1}^n J_i(z) = 0.
\eeq
Applying \eqref{betaCF} to the weights $v_i$ \eqref{nwt}, 
we can see that
the spin-$1$ currents $J_i(z)$ \eqref{Ji} are,
as collective fields of eigenvalues,  given by
\beq
\label{CFA}
J_i(z) = \frac{1}{g_s} t_i(z) - \im b \sum_{a=1}^{n-1}
( \delta_{i,a} - \delta_{i,a+1} ) \mathrm{Tr}
\, \frac{1}{z - M_a },
\eeq
where
\beq
t_i(z):= ( v_i, W'(z) ) = \sum_{a=i}^{n-1} W'_a(z) 
- \frac{1}{n} \sum_{a=1}^{n-1}
a \, W_a'(z), \qquad
i=1,2,\dotsc, n.
\eeq

The partition function of the $A_{n-1}$ quiver matrix models
obeys the $W_n = W(A_{n-1})$ constraints.
The generators of $W_n$ algebra are currents $\mathcal{W}^{(s)}(z)$
with spin $s$ ($s=2,3, \dotsc, n$). It can be constructed from
the spin $1$-currents $J_i(z) = \im \partial \varphi_i(z)$ 
by the Miura transformation \cite{FL}
\beq
: \! \det\bigl( x - \im g_s \partial \phi(z) \bigr) \! : \ 
= \ : \! \!  \prod_{1 \leq i \leq n}^{\leftarrow}
\bigl( x - g_s J_i(z) \bigr) \! : \ 
= x^n + \sum_{k=2}^{n}(-1)^k g_s^k \, \mathcal{W}^{(k)}(z) x^{n-k}.
\eeq
Here $[x, z] = -\im Q g_s$ and we use the following ordering
of product for non-commuting objects $A_i$:
\beq
\prod_{i_0 \leq i \leq i_1}^{\leftarrow} A_i:=
A_{i_1} A_{i_1-1} \dotsm A_{i_0+1} A_{i_0}.
\eeq

In particular, the current $\mathcal{W}^{(2)}(z)$
is proportional to the energy momentum tensor $T(z)$
\beq
\begin{split}
\label{AnEM}
T(z) &= - \mathcal{W}^{(2)}(z) \cr
&= \sum_{1 \leq i< j \leq n}: \partial \varphi_i(z)
\partial \varphi_j(z) : 
+ Q \sum_{i=1}^n ( n-i) \partial^2 \varphi_i(z) \cr
&= - \frac{1}{2} \sum_{i=1}^n
: \bigl( \partial \varphi_i(z) \bigr)^2: 
+ \frac{Q}{2} \sum_{i=1}^n (n-2i+1) \partial^2
\varphi_i(z) \cr
&= - \frac{1}{2} : \mathrm{Tr}\, ( \partial \phi(z) )^2: 
+ Q \langle \rho, \partial^2 \phi(z) \rangle.
\end{split}
\eeq
In \eqref{AnEM}, to go to the third line and the last,
we used the traceless condition $\sum_i \varphi_i=0$ and
the explicit form of the Weyl vector $\rho$ \eqref{WVAn}, respectively.
The central charge of this energy momentum tensor is
$c=(n-1)(1+n(n+1)Q^2)$.

The $W_n$ constraints are equivalent to the regularity of the
correlation function
\beq
\bigl\langle \! \bigl\langle 
\det( x - \im g_s \partial \phi(z) ) 
\bigr\rangle \! \bigr\rangle.
\eeq
Let us see this correlator is well-defined at finite $N_a$.
The collective field current $J_i(z)$ \eqref{CFA} has simple poles
at $z = \lambda^{(i)}_I$ and at $z = \lambda^{(i-1)}_I$.
As a function of $z$,  
\beq
\det( x - \im g_s \partial \phi(z))
= \prod_{1 \leq i< n}^{\leftarrow} ( x - g_s J_i(z) )
\eeq
has poles at $z = \lambda^{(a)}_I$ which may cause singularity
of its correlator $\langle\! \langle \det( x - \im \partial \phi(z))
\rangle \! \rangle $
at real $z$.
First, let us examine singularity at $z = \lambda^{(1)}_I$.
Note that from the form of the collective field current \eqref{CFA},
the singularities only come from the factor $(x - g_s J_2(z))(x - g_s J_1(z))$.
The other factor $(x - g_s J_n(z))\dotsm (x - g_s J_3(z))$
is independent of $\lambda^{(1)}_I$. It is not difficult to check that
near $z \rightarrow \lambda^{(1)}_I$,
\beq
\prod_{1 \leq j \leq n}^{\leftarrow}(x - g_s J_j(z))
= - g_s^2 \frac{\partial}{\partial \lambda_I^{(1)} }
\left( \prod_{3 \leq j \leq n}^{\leftarrow} ( x - g_s J_j(z))
\, 
\frac{1}{z - \lambda^{(1)}_I} \, 
\ex^{-S_{\mathrm{eff}} } \right)
\ex^{S_{\mathrm{eff}} } + O(1),
\eeq
where the effective action $S_{\mathrm{eff}}$ is defined by
\beq
\ex^{-S_{\mathrm{eff}}}:=  
\bigl( \Delta_{A_{n-1}}(\lambda))^{-b^2}
\exp\left( - \frac{\im b}{g_s} \sum_{a=1}^{n-1}
\sum_{I=1}^{N_a} W_a(\lambda^{(a)}_I ) \right).
\eeq
Similar relations hold near other points $z \rightarrow \lambda^{(a)}_I$.
Therefore, the singular part of $\det(x - \im g_s \partial \phi(z))$
is given by
\beq
\begin{split}
& \qquad \det( x - \im g_s \partial \phi(z) ) \Bigr|_{\mathrm{singular}} 
\times   \ex^{ - S_{\mathrm{eff}} }\cr
&= - g_s^2 
\sum_{a=1}^{n-1} \sum_{I=1}^{N_a}
\frac{\partial}{\partial \lambda^{(a)}_I}
\left( \prod_{i+2 \leq j \leq n}^{\leftarrow}
(x - g_s J_j(z) )\, \frac{1}{z - \lambda^{(a)}_I } \,
\prod_{1 \leq k \leq i-1}^{\leftarrow} ( x - g_s J_k(z))
\, \ex^{-S_{\mathrm{eff}} } \right).
\end{split}
\eeq
Hence we have the $W_n$ constraints:
\beq
\langle \! \langle \, \det( x - \im g_s \partial \phi (z) )
\Bigr|_{\mathrm{singular}} \rangle\! \rangle = 0.
\eeq

Therefore, we can use this correlator to define a non-commutative
Calabi-Yau threefold as follows:
\beq
u v + \langle \! \langle \det( x - \im g_s \partial \phi(z) )
\rangle \! \rangle = 0,
\eeq
and the associated non-commutative spectral curve $\Sigma$ can
be defined by
\beq \label{NCSC}
\langle \! \langle \det ( x - \im g_s \partial \phi(z) \rangle\!
\rangle = 0,
\eeq
with the non-commutativity $[x,z] = - \im Q g_s$.
These are well-defined at finite $N_a$.

Note that the correlator takes the form
\beq 
\bigl\langle \! \bigl\langle \,
\det( x - \im g_s \partial \phi(z))\,
\bigr\rangle \! \bigr\rangle
= x^n + \sum_{k=2}^n (-1)^k P_k(z) \, x^{n-k}, \qquad
P_k(z) = g_s^k \, \bigl\langle \! \bigl\langle 
\mathcal{W}^{(k)}(z)
\bigr\rangle\! \bigr\rangle.
\eeq
In principle, the explicit form of $P_k(z)$ can be determined by
examining regular part of $\det(x - \im g_s \partial \phi(z))$.
But actual calculation is tedious for general $n$. In general,
the non-commutative geometry takes the form
\beq
u v + \prod_{1 \leq i \leq n}^{\leftarrow} ( x - t_i(z))
+ \dotsm = 0.
\eeq
In the next subsection, we give the explicit form of the
spectral curve for $n=3$. 

Before going to the case of $n=3$, 
let us consider the large $N$ limit.
As in the one-matrix model case, we take $g_s \rightarrow 0$
limit with the deformed 't Hooft couplings $\tilde{S}_a:= -\im b g_s N_a$
fixed. In the large $N$, the saddle points are determined by
the stationary condition 
$\partial S_{\mathrm{eff}}/\partial \lambda^{(a)}_I=0.$
In the $g_s \rightarrow 0$ limit, 
the non-commutativity is lost and the large $N$ factorization gives
\beq
\lim_{g_s \rightarrow 0} 
\bigl\langle \! \bigl\langle \,
\det( x - \im g_s \partial \phi(z))\,
\bigr\rangle \! \bigr\rangle
= \prod_{i=1}^n ( x - y_i(z) ),
\eeq
where
\beq \label{yiz}
y_i(z) := \lim_{g_s \rightarrow 0}
\im g_s \langle \! \langle \partial \varphi_i(z) \rangle \! \rangle
=  t_i(z) + \sum_{a=1}^{n-1} ( \delta_{i,a} 
- \delta_{i,a+1} ) \, \tilde{S}_a \int \frac{\rho_a(\lambda)}{z - \lambda}
\de \lambda.
\eeq
Here the density functions $\rho_a(\lambda)$ are defined by
\beq
\rho_a(\lambda):= \lim_{N_a \rightarrow \infty} \frac{1}{N_a}
\sum_{I=1}^{N_a} \delta(\lambda - \lambda^{(a)}_I), \qquad
a=1,2,\dotsc, n-1.
\eeq
In the deformed $A_{n-1}$ quiver matrix model, 
the deformed geometry takes the form
\beq
u v + \prod_{i=1}^n ( x - y_i(z)) 
= u v + \prod_{i=1}^n ( x - t_i(z) ) + \dotsm = 0.
\eeq
Therefore, the deformed spectral curve $\Sigma$
is covered by $n$ Riemann sheets and 
the points on the $i$-th sheet can be parametrized by
$(x, z) = ( y_i(z), z)$. Hence they are fibered over the complex
$z$-plane.

Generalization to 
$D_r$ and $E_r$ cases is  straightforward 
at least in the large $N$ limit. 
Replace the 't Hooft couplings $S_a=g_s N_a$
by the deformed ones $\tilde{S}_a=-\im b g_s N_a = \epsilon_1 N_a$.
The finite non-commutative form of the spectral curve
would be fixed by requiring the consistency with the symmetry
of the partition function \eqref{partb}.

\subsection{Deformed $A_2$ quiver model and non-commutative curve}
\label{subsec:A2spectral}

For $\beta$-deformed $A_2$ quiver matrix model,
the explicit forms of the currents are given by
\beq
\begin{split}
g_s J_1(z) &= t_1(z) -\im b g_s \, \mathrm{Tr}\, \frac{1}{z- M_1}, \cr
g_s J_2(z) &= t_2(z) + \im b g_s \, \mathrm{Tr}\, \frac{1}{z-M_1}
- \im b g_s \, \mathrm{Tr}\, \frac{1}{z-M_2}, \cr
g_s J_3(z) &= t_3(z) + \im b g_s \, \mathrm{Tr}\, \frac{1}{z-M_2},
\end{split}
\eeq
where
\bea
\label{tiz}
t_1(z)
 =  \frac{1}{3} \bigl( 2 W_1'(z) + W_2'(z) \bigr),~~~~
t_3(z)
 = - \frac{1}{3} \bigr( W_1'(z) + 2 W_2'(z) \bigr),
\eea
and $t_2 = - (t_1 + t_3)$.
With some work, we can see that at finite $N_a$
(with $[x,z] = -\im Q g_s$), the regular part of the
``spectral determinant'' is given by
\beq
\label{RP3}
\begin{split}
& \prod_{1 \leq i \leq 3}^{\leftarrow}
(x - g_s J_i(z)) \Bigr|_{\mathrm{regular}} \cr
&= \prod_{1 \leq i \leq 3}^{\leftarrow} ( x - t_i(z)) \cr
&- (x - t_3(z))(-\im b g_s)
\sum_{I=1}^{N_1} \frac{W_1'(z) - W_1'(\lambda_I^{(1)})}
{z - \lambda_I^{(1)}}
- (-\im b g_s)
\sum_{J=1}^{N_2} \frac{W_2'(z) - W_2'(\lambda_J^{(2)})}
{z - \lambda_J^{(2)}} ( x - t_1(z)) \cr
& -(- \im b g_s)^2 \sum_{I=1}^{N_1}
\sum_{J=1}^{N_2} \frac{W_1'(z) - W_1'(\lambda_I^{(1)})}
{(z-\lambda_I^{(1)})(\lambda_I^{(1)} - \lambda_J^{(2)})}
 + (- \im b g_s)^2 \sum_{I=1}^{N_1}
\sum_{J=1}^{N_2} \frac{W_2'(z) - W_2'(\lambda_J^{(2)})}
{(z-\lambda_J^{(2)})(\lambda_J^{(2)} - \lambda_I^{(1)})} \cr
& + g_s^2 \sum_{I=1}^{N_1}
\frac{\partial}{\partial \lambda^{(1)}_I}
\left( \frac{t_3(z) - t_3(\lambda^{(1)})}{z - \lambda_I^{(1)}}
\ex^{-S_{\mathrm{eff}} } \right)
\ex^{S_{\mathrm{eff}} }
+ g_s^2 \sum_{J=1}^{N_2}
\frac{\partial}{\partial \lambda^{(2)}_J}
\left( \frac{t_1(z) - t_1(\lambda^{(1)})}{z - \lambda_J^{(2)}}
\ex^{-S_{\mathrm{eff}} } \right)
\ex^{S_{\mathrm{eff}} }.
\end{split}
\eeq
The last two terms in the right-handed side of \eqref{RP3}
do not contribute to the correlator.
Therefore, we have the explicit form of the finite $N_a$
spectral curve mentioned in the introduction:
\beq
\begin{split}
0&= \bigl\langle \! \bigl\langle \det( x - \im g_s \partial \phi(z))
\bigr\rangle \! \bigr\rangle \cr
&= \prod_{1 \leq i \leq 3}^{\leftarrow} ( x - t_i(z))
- \bigl( x - t_3(z) \bigr) f_1(z)
- f_2(z) \bigl( x - t_1(z) \bigr)  - g_1(z) + g_2(z),
\end{split}
\eeq
where
\beq
\label{fa}
f_a(z):= \left\langle \!\!\! \left\langle
- \im b g_s \sum_{I=1}^{N_a} \frac{W_a'(z) - W_a'(\lambda_I^{(a)})}
{z - \lambda_I^{(a)}}
\right\rangle \!\!\! \right\rangle,
\eeq
\beq
\label{ga}
g_a(z):= \left\langle \!\!\! \left\langle
(- \im b g_s)^2 \sum_{I=1}^{N_a}
\sum_{J=1}^{N_{\bar{a}}} \frac{W_a'(z) - W_a'(\lambda_I^{(a)})}
{(z-\lambda_I^{(a)})(\lambda_I^{(a)} - \lambda_J^{(\bar{a})})}
\right\rangle \!\!\! \right\rangle,
\eeq
with $\bar{a} = 3 - a$.
At the commutative point $[x, z]=0$ $(b=\im)$, the
algebraic curve of this form (in the large $N$ limit) is
already known \cite{DV2002b, sek, Hofman, Lazaroiu, KLLR, CT}.

In the large $N$ limit with general $b$, spectral curve factorizes into
\beq
\label{LNspc}
\begin{split}
0&=\prod_{i=1}^3 ( x - t_i(z)) - f_1(z)(x-t_3(z))
-f_2(z)(x-t_1(z)) - g_1(z) + g_2(z) \cr
&= \prod_{i=1}^3 ( x - y_i(z) ),
\end{split}
\eeq
where
\beq
\label{A2yi}
y_1(z) = t_1(z) + \tilde{S}_1 \omega_1(z), ~~
y_2(z) = t_2(z) - \tilde{S}_1 \omega_1(z) + \tilde{S}_2 \omega_2(z), ~~
y_3(z) = t_3(z) - \tilde{S}_2 \omega_2(z).
\eeq
Here $\tilde{S}_a = - \im b g_s N_a$ and
\beq
\omega_a(z) = \lim_{N_a \rightarrow \infty}
\frac{1}{N_a} \left\langle \!\!\! \left\langle
\mathrm{Tr}\, \frac{1}{z - M_a} \right\rangle\!\!\!\right\rangle
= \int \frac{\rho_a(\lambda)}{z-\lambda} \de \lambda.
\eeq

\section{Dijkgraaf-Vafa's proposal: multi-Penner potential}
  In \cite{AGT, Wyllard}, it was pointed out that the correlation function of
  $A_{n-1}$ Toda field theory (Liouville theory for $n=2$) in two dimensions can be associated
  with Nekrasov's partition function \cite{Nekrasov, NO} of $SU(n)$ quiver gauge theory.
  For $SU(n)$ gauge theory with $2n$ hypermultiplets
  its Nekrasov partition function is identified
  with the (chiral) four point function of the Toda theory on a sphere:
    \bea
    \left< V_{\hat{\mu_0}}(z_0) V_{\hat{\mu}_1}(z_1) V_{\hat{\mu}_2}(z_2) V_{\hat{\mu}_3}(z_3)  \right>,
    \label{correlationToda}
    \eea
  under the identification (\ref{NekrasovCFT})
  between the deformation parameters and the parameter $b$ 
appearing in the central charge \eqref{ccADE} of the Toda theory.
 
  The positions of the vertex operators correspond to
 punctures of the sphere.
  Three of them can be chosen at $z = 0, 1, \infty$.
  As we will see later, this sphere can be identified with the one on which $n$ M5-branes wrap.
  The vectors $\hat{\mu}_p$ in the dual space $\mathfrak{h}^*$ of the Cartan subalgebra
  are expanded as $\hat{\mu}_p = \sum_a \hat{\mu}_{p,a} \Lambda^a$.
  In general, it is possible to consider many types of punctures in the sphere on which M5-branes wrap.
  This variety corresponds to the 
flavor symmetries of the gauge theory \cite{Gaiotto}.
  In the Toda theory, this corresponds to choice of the vectors $\hat{\mu}_p$.
  The vertex operator $V_{\hat{\mu}}(z)$ corresponding to the ``simple" puncture
  associated with the $U(1)$ flavor symmetry is \cite{Wyllard}
    \bea
    \hat{\mu}
     \propto
           \Lambda^1
           ~~
           {\rm or}~~
           \Lambda^{n - 1}
           \label{U1puncture}
    \eea
  where $\Lambda^1$ ($\Lambda^{n - 1}$) is the highest weight of the (anti-)fundamental representation.
  The other types of punctures correspond to more generic choices of $\hat{\mu}$.
  The $SU(n)$ gauge theory with $2n$ hypermultiplets has $SU(n)^2 \times U(1)^2$ flavor symmetry
  as a subgroup of $U(2n)$.
  Therefore, for this gauge theory, the corresponding correlation function is such that
  two vertex operators are of ``simple" type like (\ref{U1puncture}) while the other two are of generic ``$SU(n)$" type.
 
  As discussed in \cite{DV},
  by using the relation between $A_{n-1}$ Toda field theory and the quiver matrix model \cite{MMM, KMMMP},
  we reach the matrix model which describes $\CN=2$ $SU(n)$ quiver gauge theory.
  The prescription introduced in \cite{DV} is that the original matrix model action is set to zero.
  Under zero action, we consider the correlation functions of the vertex operators considered above.
  In the correlation function, the vertex operator can be written in terms of the matrices by using (\ref{betaCF}) as
    \bea
\label{VOdet}
    V_{\hat{\mu}}(z)
     =     : \ex^{\im \left< \hat{\mu}, \phi(z) \right>} :
     \rightarrow   
           \prod_{a = 1}^{n - 1} \det (z - M_a)^{- \im b \left( \hat{\mu}, \alpha_a \right)}
    \eea

With zero action ($W_a(z)=0$), we define the chiral four-point
correlation function which corresponds to
  $SU(n)$ gauge theory with $2n$ hypermultiplets by
\begin{equation}
\left\langle
: \ex^{\im \langle \hat{\mu}_0, \phi(q_0) \rangle} ::
\prod_{p=1}^3 \ex^{\im \langle \hat{\mu}_p , \phi(q_p) \rangle}:
\right\rangle 
\equiv \langle 0 | 
: \ex^{\im \langle \hat{\mu}_0, \phi(q_0) \rangle} ::
\prod_{p=1}^3 \ex^{\im \langle \hat{\mu}_p , \phi(q_p) \rangle}:
Q_1^{N_1} Q_2^{N_2} \dotsm Q_{n-1}^{N_{n-1}} | 0 \rangle,
\end{equation}
where $Q_a = \int \de \lambda : \ex^{b \phi_a(\lambda)}:$.
For later convenience, we set $\mu_p:= g_s \hat{\mu}_p$
$(p=0,1,2,3)$.
The momentum conservation condition is required
\bea
\label{momentumconservation}
\mu_0 + \sum_{p=1}^3 \mu_p + 
\sum_{a=1}^{n-1} \tilde{S}_a \alpha_a = 0.
\eea
Using this four-point function, 
we define the partition function of the
deformed $A_{n-1}$ 
quiver matrix model by
\begin{equation}
\begin{split}
Z&:= \lim_{q_0 \rightarrow \infty}
q_0^{(\hat{\mu}_0, \hat{\mu}_0)}
\left\langle
: \ex^{\im \langle \hat{\mu}_0, \phi(q_0) \rangle} ::
\prod_{p=1}^3 \ex^{\im \langle \hat{\mu}_p , \phi(q_p) \rangle}:
\right\rangle \cr
&= \int \prod_{a=1}^{n-1}
\left\{ \prod_{I=1}^{N_a} \de \lambda^{(a)}_I \right\}
\bigl(\Delta_{A_{n-1}}(\lambda) \bigr)^{-b^2}
\exp\left( - \frac{\im b}{g_s} \sum_{a=1}^{n-1}
\sum_{I=1}^{N_a} W_a(\lambda^{(a)}_I) \right),
\end{split}
\end{equation}
where the determinantal form of the vertex operators \eqref{VOdet}
at punctures $q_p$
is converted into a set of new matrix model potentials $W_a(z)$
with multi-Penner type interaction:
    \bea \label{multPenWa}
    W_a(z)
     =     \sum_{p = 1}^3 \left( \mu_p, \alpha_a \right)
           \log (q_p - z).
           \label{matrixactionn}
    \eea
Later we will set $q_1 = 0$, $q_2 = 1$ and $q_3 = q$.

We denote the components of $\mu_p$ $(p=0,1,2,3)$
in the fundamental weight basis by $\mu_{p,a}$:
$\mu_p = \sum_a \mu_{p,a} \Lambda^a$.
As explained above, we let the operators at 
$z = q_2, q_3$ be of the simple type.
That is, we set
    \bea
    \mu_2
     =     \mu_{U(1)_1} \Lambda^1,
           ~~~~
    \mu_3
     =     \mu_{U(1)_2} \Lambda^{n-1}.
           \label{U1punctures}
    \eea
  The operators at $z = q_1, \infty$ are of the generic type.

\subsection{$A_2$ quiver matrix model corresponding to $SU(3)$ gauge theory}
\label{subsec:A2gauge}
  In this subsection, we consider $A_2$ quiver matrix model with the action (\ref{matrixactionn}).
  As seen above, the Nekrasov partition function for $\CN=2$ supersymmetric $SU(3)$ gauge theory with six massive flavors
  is expected to be described by $A_2$ quiver matrix model with the action (\ref{matrixactionn}) with $n=3$.
 
  We have explicitly seen the spectral curve of $A_2$ quiver matrix model in the large $N_a$ limit
  in the subsection \ref{subsec:A2spectral}.
  For convenience, we rewrite \eqref{LNspc} in the following form:
    \bea
    x^3
     =     p(z) x + q(z),
           \label{A2curve}
    \eea
  where
    \bea
    p(z)
    &=&    t_1(z)^2 + t_3(z)^2 + t_1(z) t_3(z) + f_1(z) + f_2(z),
           \nonumber \\
    q(z)
    &=&  - t_1(z) t_3(z) (t_1(z) + t_3(z)) -f_1(z) t_3(z) - f_2(z) t_1(z) + g_1(z) - g_2(z).
    \eea
  By substituting the action \eqref{matrixactionn}
and (\ref{U1punctures}), the explicit forms of $t_i(z)$ \eqref{tiz} are given by
    \bea
    t_1(z)
    &=&    \frac{1}{3} \sum_{p = 1}^3 \frac{\left( \mu_p, 2 \alpha_1 + \alpha_2 \right) }{z - q_p}
     =     \frac{1}{3} \left( \frac{ 2\mu_{1,1} + \mu_{1,2} }{z}
         + \frac{2\mu_{U(1)_1}}{z - 1} + \frac{\mu_{U(1)_2}}{z - q} \right),
           \nonumber \\
    t_3(z)
    &=&  - \frac{1}{3} \sum_{p = 1}^3 \frac{\left( \mu_p, \alpha_1 + 2\alpha_2 \right) }{z - q_p}
     =   - \frac{1}{3} \left( \frac{\mu_{1,1} + 2 \mu_{1,2}}{z}
         + \frac{\mu_{U(1)_1}}{z - 1} + \frac{2 \mu_{U(1)_2}}{z - q} \right),
    \eea
  and $t_2(z) = - t_1(z) - t_3(z)$. Here $\mu_{1,a}$ are components of $\mu_1$ in the fundamental weight basis:
  $\mu_1 = \mu_{1,1} \Lambda^1 + \mu_{1, 2} \Lambda^2$.

  The form of $f_a$ \eqref{fa} and that of $g_a$ \eqref{ga}
are
    \bea
    f_a(z)
     =     \sum_{p = 1}^3 \frac{f_p^{(a)}}{z - q_p},
           ~~~
    g_a(z)
     =     \sum_{p = 1}^3 \frac{g_p^{(a)}}{z - q_p},
    \eea
  where $f_p^{(a)}$ and $g_p^{(a)}$ are constants.
  As seen above, the coordinate $z$ parametrizes the sphere.
  The spectral curve is a triple cover of this sphere
  and each value of $x$ at each sheet is given by $y_i(z)$.
 
  We consider a one-form $\lambda_m:= y_i(z) \, \de z$ $(i=1,2,3)$
with $y_i(z)$ given by \eqref{A2yi}.
Later on
this will be identified with the Seiberg-Witten one-form
on the $i$-th sheet.
  From the form of the spectral curve,
we can see that the one-form has simple poles 
at $z = q_p$ and at $z = \infty$.

  The residues of $y_i(z) \de z$ $(i=1,2,3)$ at the simple punctures $z=1$ and $z=q$ are given by
    \bea
    \mathrm{res}_{z=1}\bigl( y_i(z) \de z \bigr)
    &=&    \frac{1}{3} \left( 2 \mu_{U(1)_1}, -\mu_{U(1)_1}, - \mu_{U(1)_1} \right),
           \label{z=1residues} \\
    \mathrm{res}_{z=q} \bigl( y_i(z) \de z \bigr)
    &=&    \frac{1}{3} \left( \mu_{U(1)_2}, \mu_{U(1)_2}, - 2 \mu_{U(1)_2} \right).
           \label{z=qresidues}
    \eea
  These will be compared with the mass parameter associated with the $U(1)$ flavor symmetry in the gauge theory.
  The residues at $z = 0$ are of generic type:
\bea
\mathrm{res}_{z=0}\bigl( y_i(z) \de z \bigr)
=\frac{1}{3} \left(
\, 2 \mu_{1,1} + \mu_{1,2}\, , \, -\mu_{1,1} + \mu_{1,2}\, , \,
- \mu_{1,1} - 2 \mu_{1,2} \, \right).
\label{z=0residues}
\eea
  This takes the same form as the mass parameters associated with the $SU(3)$ flavor symmetry,
  which will be analyzed in the next section.
 
Using the momentum conservation \eqref{momentumconservation},
the residues at $z = \infty$ are found to be
    \bea
\label{z=inftyresidues}
    \mathrm{res}_{z=\infty}\bigl( y_i(z) \de z \bigr)
     =    \frac{1}{3} \left( 2 \mu_{0,1} + \mu_{0,2}, -\mu_{0,1} + \mu_{0,2}, - \mu_{0,1} - 2 \mu_{0,2} \right).
    \eea
  This also takes the same form as the $SU(3)$ mass parameters.

\section{$\CN=2$ gauge theory and Seiberg-Witten curve}
\label{sec:gauge}
  We consider $\CN=2$ $SU(n)$ gauge theory with $N_f=2n$.
  The low energy effective theory (and the quantum Coulomb moduli space) of $\CN=2$ supersymmetric gauge theory
  is described by the Seiberg-Witten curve
  and the meromorphic one-form called Seiberg-Witten one-form \cite{SeibergWitten, KLYT, AF, HO, APS}.
  We will see that the Seiberg-Witten curve of this theory enjoys the same properties as the spectral curve
  considered above for the $SU(n)$ ($A_{n-1}$) case.

\subsection{$SU(3)$ gauge theory with $N_f = 6$}

We first consider the $SU(3)$ gauge theory with six massive hypermultiplets.
  This theory has an exactly marginal coupling 
which is the microscopic gauge coupling constant:
    \bea
    \tau_{\mathrm{UV}}
     =     \frac{\theta_{{\rm UV}}}{\pi} + \frac{8 \pi \im}{g_{{\rm UV}}^2}.
           \label{gaugecoupling}
    \eea
  The type IIA brane construction and its M-theory lift lead to
  the Seiberg-Witten curve which is a hypersurface 
in 2-complex dimensional space $(t, v)$ \cite{Witten,Gaiotto}
    \bea
    & &    (v - m_1)(v - m_2)(v - m_3)t^2 - (1 + q_{{\rm UV}}) (v^3 + P v^2 + Q v + R)t
           \nonumber \\
    & &    ~~~~~~~~~~~~~~~~~~~~~~~~~~~~~~~~~~~~~
         + q_{{\rm UV}} (v - m_4)(v - m_5)(v - m_6)
     =     0,
           \label{SWcurveSU(3)M}
    \eea
  where $m_i$ ($i = 1, \ldots, 6$) are the mass parameters of the hypermultiplets
  and $q_{{\rm UV}} = e^{\pi \im \tau_{{\rm UV}}}$.
  We can see that the coordinate $v$ has  mass dimension one
  and therefore the constants $P$, $Q$ and $R$ have mass dimension one, two and three respectively.
  In principle, the constants depend on the mass parameters
  and the Coulomb moduli parameters $u^{(2)}$ and $u^{(3)}$
  which correspond to $\langle \Tr \phi^2 \rangle$ and $\langle \Tr \phi^3 \rangle$ at weak coupling.
  The dimensional analysis and the regularity constraint in the massless limit show
  that $Q$ and $R$ are, respectively, linear in $u^{(2)}$ 
and $u^{(3)}$ and also include the terms $m^2$ and $m^3$.
  Also, $P$ depends only on the mass parameters.
  The curve is translated into the following form
    \bea
    (t - 1)(t - q_{{\rm UV}}) v^3
     =     M_2 (t) v^2 + V^{(2)}_2(t) v + V_2^{(3)}(t),
    \eea
  where $M_2$, $V_2^{(2)}$ and $V_2^{(3)}$ are degree two polynomials in $t$:
\begin{equation}
\begin{split}
M_2(t) &= \left( \sum_{i=1}^3 m_i \right) t^2 + P ( 1 + q_{\mathrm{UV}}) \, t
+ q_{\mathrm{UV}} \left( \sum_{j=4}^6 m_j \right), \cr
V_2^{(2)}(t) &= - \left( \sum_{1 \leq i<j \leq 3} m_i m_j \right) t^2
+ Q ( 1 + q_{\mathrm{UV}} ) t - q_{\mathrm{UV}}
\left( \sum_{4 \leq i<j \leq 6} m_i m_j \right), \cr
V_2^{(3)}(t) &= m_1 m_2 m_3 \, t^2 + R ( 1 + q_{\mathrm{UV}} ) \, t + 
m_4 m_5 m_6 \, q_{\mathrm{UV}}.
\end{split}
\end{equation}

  As in \cite{Gaiotto}, by shifting the coordinate $v$ appropriately, we can eliminate the quadratic term in $v$.
  Then, by changing the coordinate $v = x t$, we obtain \cite{Gaiotto}
    \bea
    x^3
     =     \frac{P^{(2)}_4(t)}{(t (t - 1) (t - q_{{\rm UV}}))^2} x
         + \frac{P^{(3)}_6(t)}{(t (t - 1) (t - q_{{\rm UV}}))^3},
           \label{SWcurveSU(3)}
    \eea
  where $P^{(2)}_4(t)$ and $P^{(3)}_6(t)$ are the degree 4 and 6 polynomials respectively
  and can be written in terms of $M_2$, $V_2^{(2)}$ and $V_2^{(3)}$
    \bea
    P^{(2)}_4(t)
    &=&    \frac{M_2(t)^2}{3} + V_2^{(2)}(t) (t - 1)(t - q_{{\rm UV}}),
           \nonumber \\
    P^{(3)}_6(t)
    &=&    \frac{2 M_2(t)^3}{27} + \frac{1}{3} M_2(t) V_2^{(2)}(t) (t - 1)(t - q_{{\rm UV}})
         + V_2^{(3)}(t) (t - 1)^2(t - q_{{\rm UV}})^2.
    \eea
 
  In this coordinate, the Seiberg-Witten one-form is $\lambda_{{\rm SW}} = x \, \de t$.
  We denote by $t$ a coordinate in the sphere on which three M5-branes wrap.
  Therefore, $(t, x)$ are local coordinates in the cotangent bundle of this sphere.
  We identify the sphere with the one that appeared in the quiver matrix model and Toda theory.
  This implies $q = q_{{\rm UV}}$.
 
  From the expression (\ref{SWcurveSU(3)}),
we can see that the Seiberg-Witten one-form has 
poles at $t = 0, 1, q_{{\rm UV}}$ and $\infty$.
  As found in \cite{Gaiotto} and we will see below,
the structure of those poles corresponds to 
the flavor symmetry of the quiver gauge theory,
because the residues of the Seiberg-Witten one-form 
are the mass parameters
associated with the flavor symmetries.

The residues of the Seiberg-Witten one-form $x \, \de t$ 
at $t=1$ and $t=q_{\mathrm{UV}}$ are given by
\begin{equation}
\label{t=1residues}
\begin{split}
\mathrm{res}_{t=1}\bigl( x \, \de t \bigr)
&= \frac{M_2(1)}{3(1-q_{\mathrm{UV}})}
\left( 2, -1, -1 \right), \cr
\mathrm{res}_{t=q_{\mathrm{UV}}} \bigl( x\, \de t \bigr)
&= \frac{M_2(q_{\mathrm{UV}})}{3 q_{\mathrm{UV}} ( 1 - q_{\mathrm{UV}})}
\left( 1, 1, -2 \right),
\end{split}
\end{equation}
where each value denotes the residue at the $i$-th sheet of the Riemann surface.

The residues at $t=0$ are found to be
\bea
\mathrm{res}_{t=0} \bigl( x\, \de t \bigr)
= \frac{1}{3} ( 2 m_4 - m_5- m_6, 
- m_4 + 2 m_5 - m_6 ,
- m_4 - m_5 + 2 m_6).
\label{t=0residues}
\eea

  The flavor symmetry $U(1) \times SU(3)$ associated 
with the punctures at $t = 1, \infty$ are the symmetry of
  three hypermultiplets whose masses are $m_i$ ($i = 1,2,3$).
  Since the mass associated with $U(1)$ is $\sum_{i=1}^3 m_i/3$, 
the residue at $t = 1$ should be of the form
    \bea
    \frac{1}{3} \left( 2 \sum_{i = 1}^3m_i, - \sum_{i = 1}^3m_i, - \sum_{i = 1}^3m_i \right).
    \eea
  Also, the residue at $t = q_{{\rm UV}}$ should be of the form
\bea
\frac{1}{3} \left( \sum_{j=4}^6 m_j, 
\sum_{j=4}^6 m_j, - 2 \sum_{j=4}^6 m_j \right).
\eea
  These determine $M_2(t)$ completely and lead to
    \bea
    P
     =   - \frac{q_{{\rm UV}}}{1 + q_{{\rm UV}}} \sum_{i = 1}^6 m_i.
    \eea
  From this, we can compute the residues at $t = \infty$ as
    \bea
\label{t=inftyresidues}
    \frac{1}{3} \left(
- 2m_1 + m_2 + m_3, 
~m_1 - 2m_2 + m_3, 
~m_1 + m_2 - 2m_3 \right),
    \eea
  which corresponds to the Cartan part of $SU(3)$.
  At $t = 0$, we also find the similar form as above.
   From the residue analysis alone, we cannot determine 
the constants $Q$ and $R$ in (\ref{SWcurveSU(3)M}).

\subsubsection{Matching of the matrix model parameters}

Obviously, the coordinate $t$ is identified with $z$ and $x \, \de t$
on the $i$-th sheet 
with $y_i \, \de z$.
Comparing the residues \eqref{z=1residues}, \eqref{z=qresidues} with
\eqref{t=1residues}, we find that
the parameters in $A_2$ quiver matrix model must be chosen as
\bea
\mu_{U(1)_1} = \sum_{i=1}^3 m_i, \qquad
\mu_{U(1)_2} = \sum_{j=4}^6 m_j,
\eea
in order to yield the Seiberg-Witten curve of the gauge theory.
Also, by comparing \eqref{z=inftyresidues} with \eqref{t=inftyresidues},
and \eqref{z=0residues} with \eqref{t=0residues}, we have
\bea
\mu_{0,1} = -m_1 + m_2, \qquad \mu_{0,2} = -m_2 + m_3, \qquad
\mu_{1,1} = m_4 - m_5, \qquad
\mu_{1,2} = m_5 - m_6.
\eea
To summarize, we have determined the ``weights'' of the vertex
operators $V_{(\mu_p/g_s)}(q_p)$ as follows:
\begin{align}
\mu_0 &= (-m_1 + m_2) \Lambda^1 + ( -m_2 + m_3)\Lambda^2,&
\mu_1 &= (m_4 - m_5) \Lambda^1 + (m_5 - m_6) \Lambda^2, \cr
\mu_2 &= ( m_1 + m_2 + m_3) \Lambda^1,&
\mu_3 &= ( m_4 + m_5 + m_6) \Lambda^2.
\end{align}

The matrix model potential $W_a(z)$ with the multi-log interaction
are finally fixed as
(up to constants)
\begin{equation}
\begin{split}
W_1(z) &= ( m_4 - m_5) \log z 
+ \left( \sum_{i=1}^3 m_i \right) \log( 1 - z), \cr
W_2(z) &= ( m_5 - m_6) \log z
+ \left( \sum_{j=4}^6 m_j \right) \log ( q_{\mathrm{UV}} - z).
\end{split}
\end{equation}

  From the momentum conservation \eqref{momentumconservation},
the deformed 't Hooft couplings $\tilde{S}_a = -\im b g_s N_a$
are related to the mass parameters of the gauge theory as follows:
\beq
\tilde{S}_1 = - m_2 - m_3 - m_4, \qquad
\tilde{S}_2 = - m_3 - m_4 - m_5.
\eeq


\subsection{Isomorphism between the matrix model curve and the Seiberg-Witten curve for general $n$} 

It is straightforward to generalize the analysis for $n=3$ to general $n$.
The spectral curve \eqref{NCSC} of the $A_{n-1}$ quiver matrix model
has the form
\beq \label{SCMMn}
x^n = \sum_{k=2}^{n} (-1)^{k-1} P_k(z) x^{n-k}, 
\eeq
where
\beq
P_k(z) = g_s^k \, \bigl\langle \! \bigl\langle 
\mathcal{W}^{(k)}(z)
\bigr\rangle\! \bigr\rangle.
\eeq
With the choice of the multi-Penner potential \eqref{multPenWa},
$P_k(z)$ have the following form of the singularities:
\beq
P_k(z) = \frac{Q_k(z)}{(z(z-1)(z-q))^k},
\eeq
for some polynomials $Q_k(z)$ in $z$.

On the other hand, 
The Seiberg-Witten curve \cite{Witten,Gaiotto} for the
$SU(n)$ gauge theory with $N_f=2n$ massive hypermultiplets is given by
\beq
\left( \prod_{i=1}^{n} ( v - m_i ) \right) t^2 
- (1 + q_{\mathrm{UV}} ) \left( v^n + P v^{n-1} + \sum_{\ell=2}^{n} Q_{\ell} v^{n-\ell} 
\right) t 
+ q_{\mathrm{UV}} \prod_{j=1}^n  ( v - \widetilde{m}_j ) = 0.
\eeq
The mass parameters of the hypermultiplets 
are denoted by $m_i$ $(i=1,2,\dotsc, 2n)$, and here
for simplicity, we have set $\widetilde{m}_j:= m_{n+j}$ 
$(j=1,2,\dotsc, n)$. 
Also $q_{\mathrm{UV}} = \ex^{\pi \im \tau_{\mathrm{UV}}}$.
This curve can be rewritten as
\beq \label{SWn2}
(t - 1 )( t - q_{\mathrm{UV}}) v^n
= M_2(t) v^{n-1} + \sum_{\ell=2}^{n} V_2^{(\ell)}(t) v^{n-\ell},
\eeq
where
\beq \label{M2t}
M_2(t) = \sigma_1(m) \, t^2 
+ (1 + q_{\mathrm{UV}}) P \, t  + q_{\mathrm{UV}}\,  
\sigma_1( \widetilde{m}) ,
\eeq
\beq
V_2^{(\ell)}(t)
= (-1)^{\ell-1} \sigma_{\ell}(m) \, t^2 + (1 + q_{\mathrm{UV}}) Q_{\ell} \, t
+ (-1)^{\ell-1} \sigma_{\ell}(\widetilde{m}).
\eeq
Here $\sigma_{\ell}(m)$ (resp. $\sigma_{\ell}(\widetilde{m})$)
is the $\ell$-th elementary symmetric polynomial of $\{ m_1, \dotsm, m_n \}$
(resp. of $\{ \widetilde{m}_1, \dotsm, \widetilde{m}_n \}$).

By a change of variable
\beq
v = x t + \frac{M_2(t)}{n(t-1)(t- q_{\mathrm{UV}})},
\eeq
the Seiberg-Witten curve \eqref{SWn2} turns into the Gaiotto form:
\beq \label{SCGTn}
\begin{split}
x^n &= 
\frac{P_4^{(2)}(t) \, x^{n-2}}
{(t(t-1)(t-q_{\mathrm{UV}}))^2} 
+ \frac{P_6^{(3)}(t) \, x^{n-3} }
{(t(t-1)(t-q_{\mathrm{UV}}))^3} 
+ \dotsm
+ \frac{P_{2n}^{(n)}(t) }{(t(t-1)(t-q_{\mathrm{UV}}))^{n}} \cr
&= \sum_{k=2}^n \frac{P_{2k}^{(k)}(t)}{( t(t-1)(t-q_{\mathrm{UV}}))^k}
x^{n-k},
\end{split}
\eeq
where $P_{2k}^{(k)}(t)$ are the following degree $2k$ polynomials in $t$:
\beq
\begin{split}
P_{2k}^{(k)}(t)
&= \frac{n! (k-1)}{k! (n-k)!} \left( \frac{M_2(t)}{n} \right)^k \cr
&+ \sum_{\ell=2}^{k}
\frac{(n-\ell)!}{(n-k)! (k-\ell)!}
\left( \frac{M_2(t)}{n} \right)^{k-\ell}
V_2^{(\ell)}(t) \, \Bigl( (t-1)( t - q_{\mathrm{UV}}) \Bigr)^{\ell-1}.
\end{split}
\eeq
By evaluating the residue at $t=1$ and at $t=q$ as in the case of $n=3$,
the constant $P$, which enters in $M_2(t)$ \eqref{M2t}, is fixed as
\beq
P = - \frac{q_{\mathrm{UV}}}{1 + q_{\mathrm{UV}}}
\sum_{i=1}^n \left( m_i + \widetilde{m}_i \right).
\eeq

Now, we can clearly see the similarity between the matrix model curve \eqref{SCMMn} 
in the planar limit and the Seiberg-Witten curve \eqref{SCGTn}.

Let us compare the residues of one-forms $y_i(z) \, \de z$ and $x \, \de t$.
In the $A_{n-1}$ quiver matrix model, 
the curve \eqref{SCMMn} in the planar limit
factorizes into
\beq
\prod_{i=1}^n ( x - y_i(z) ) =0,
\eeq
where $y_i(z)$ is defined by \eqref{yiz}.
The residues of $y_i(z) \de z$ ($i=1,2,\dotsc, n$) at the simple punctures
$z=1$ and $z=q$ are given by
\beq
\mathrm{res}_{z=1} \bigl( y_i(z) \de z \bigr)
= \frac{\mu_{U(1)_1}}{n} \Bigl( n-1, \underbrace{-1,\dotsm, -1}_{n-1} \Bigr),
\eeq
\beq
\mathrm{res}_{z=q} \bigl( y_i(z) \de z \bigr)
= \frac{\mu_{U(1)_2}}{n} \Bigl( \underbrace{1,\dotsm, 1}_{n-1}, -(n-1) \Bigr).
\eeq
The residues at generic punctures $z=0$ and $z=\infty$ are found to be
\beq
\mathrm{res}_{z=0} \bigl( y_i(z) \de z \bigr)
= \sum_{a=1}^{n-1} ( \delta_{i,a} - \delta_{i,a+1} ) \, ( \mu_1, \Lambda^a ),
\eeq
\beq
\mathrm{res}_{z=\infty} \bigl( y_i(z) \de z \bigr)
= \sum_{a=1}^{n-1} ( \delta_{i,a} - \delta_{i,a+1} ) \, ( \mu_0, \Lambda^a ),
\eeq
for $i=1,2,\dotsc, n$.
While the residues of the Seiberg-Witten one-form $x \, \de t$ at $t=1$ and
$t=q_{\mathrm{UV}}$ are given by
\beq
\begin{split}
\mathrm{res}_{t=1}(x\, \de t)
&= \frac{M_2(1)}{n(1-q_{\mathrm{UV}})} \Bigl( n-1, 
\underbrace{-1, \dotsm, -1}_{n-1} \Bigr)
=\frac{1}{n} \sigma_1(m) \Bigl( n-1, 
\underbrace{-1, \dotsm, -1}_{n-1} \Bigr), \cr
\mathrm{res}_{t=q_{\mathrm{UV}}}(x\, \de t)
&= \frac{M_2(q_{\mathrm{UV}})}{n \, q_{\mathrm{UV}} ( 1 - q_{\mathrm{UV}})}
\Bigl(\underbrace{1,\dotsm, 1}_{n-1}, -(n-1) \Bigr)
=\frac{1}{n} \sigma_1(\widetilde{m})
\Bigl(\underbrace{1,\dotsm, 1}_{n-1}, -(n-1) \Bigr),
\end{split}
\eeq
where each value denotes the residue at the $i$-th sheet of the Riemann surface.
The residues at $t=0$ and at $t=\infty$ on 
the $i$-th sheet ($i=1,2,\dotsc, n$) are given by
\beq
\begin{split}
\mathrm{res}_{t=0}( x\, \de t) 
&= \widetilde{m}_i - \frac{1}{n} \left( \sum_{j=1}^n \widetilde{m}_j \right), \cr
\mathrm{res}_{t=\infty}(x\, \de t)
&= \frac{1}{n} \left( \sum_{j=1}^n m_j\right)  - m_i.
\end{split}
\eeq

For general $n$, these residues match if 
the weights of the vertex operators are identified 
with the mass parameters of the gauge theory as follows
\beq
\mu_0 = \sum_{a=1}^{n-1} ( - m_a + m_{a+1}) \Lambda^a, \qquad
\mu_1 = \sum_{a=1}^{n-1} ( \widetilde{m}_a - \widetilde{m}_{a+1}) \Lambda^a,
\eeq
\beq
\mu_2 = \left( \sum_{i=1}^n m_i \right)
\Lambda^1, \qquad
\mu_3 = \left( \sum_{i=1}^n \widetilde{m}_i \right)
\Lambda^{n-1}.
\eeq

The matrix model potentials $W_a(z)$ ($a=1,2,\dotsc, n-1$) are fixed as
\beq
W_a(z) = ( \widetilde{m}_a - \widetilde{m}_{a+1})
\log z + \delta_{a,1} \left( \sum_{i=1}^n m_i \right) \log (1-z)
+ \delta_{a,n-1} \left( \sum_{i=1}^n \widetilde{m}_i \right) \log ( q_{\mathrm{UV}} - z).
\eeq
With this choice of the multi-log potentials, 
the $A_{n-1}$ quiver matrix model curve \eqref{SCMMn} in the planar limit 
coincides with the $SU(n)$ Seiberg-Witten curve \eqref{SCGTn}
with $N_f=2n$ massive hypermultiplets.

 It is rather simple to provide the counting of the 0d-parameters and 4d-parameters
and the matching of the numbers of parameters. With the rule concerning with
 the types (simple versus generic) of the punctures \cite{Wyllard} explained before, the four of
 the vertex operators contain $2(n-1) + 2 =2n$ parameters. This matches the number
of mass parameters in the gauge theory side.  As for the number of the Coulomb moduli,
let us first note that, in the planar limit, the $A_{n-1}$ quiver matrix model develops 
an array(or ladder) of $n-1$ spieces of two-cut eigenvalue distributions. 
The Riemann surface developed is genus $n-1$ and is depicted as a 
linear array of $n$ $P_{1}$s with adjacent two (say, $P_{1}^{(i)}$ and $P_{1}^{(i+1)}$)
connected by two pipes consisting of the $i$-th eigenvalue distribution. 
Each of the $n-1$ Coulomb moduli is obtained simply by integrating the matrix model differential
over one of the two pipes and of course represent the filling fraction of 
the $i$-th eigenvalue distribution. This is an obvious generalization of $n=1$ case
noted in \cite{DV}.

\section*{Acknowledgements}
We thank Kazuo Hosomichi for several discussions on this subject.
We also thank Shigenori Seki and Masato Nishida for earlier discussions
on the quiver matrix model and Tohru Eguchi and Masato Taki for comments on related issues.
  The research of H.~I.~ and T.~O.~
is supported in part by the Grant-in-Aid for Scientific Research (2054278)
  from the Ministry of Education, Science and Culture, Japan.
  K.~M.~ is supported in part by JSPS Research Fellowships for Young Scientists.
  H.~I.~ acknowledges the hospitality of YITP extended to him
  during the period of the workshop ``Branes, Strings and Black Holes."




\end{document}